\documentclass[aps,prd,onecolumn,a4paper,tightenlines,notitlepage,eqsecnum,11pt]{revtex4-2}
\usepackage{bbm}

\usepackage{mathrsfs}
\usepackage{bm}
\usepackage{amsfonts}
\usepackage{amsmath}
\usepackage{amssymb}
\usepackage{slashed}
\usepackage{pstricks}
\usepackage{color}

\usepackage{graphicx}

\usepackage[colorlinks=true,linkcolor=blue,citecolor=blue,breaklinks,linktoc=page,urlcolor=blue]{hyperref}

\begin{document}
\title{Novel formulation of Hamilton-Jacobi equation for higher derivative theory and quantum mechanical correspondence}

\author{Zhi-Qiang Guo}\email{gzhq@pku.edu.cn}\affiliation{Qianzhai 142, City of Linzhou, Province of Henan, China}

\begin{abstract}
For higher derivative theories, using the approach of Caratheodory's equivalent Lagrangian, we show that there exist novel formulations of Hamilton-Jacobi equations, which are different from the formulations derived from Hamilton's canonical approach. The quantum mechanical correspondences of these novel Hamilton-Jacobi equations lead to nonlinear quantum mechanics, which seem being able to avoid the unbounded negative energy problem in the quantum mechanics of higher derivative theories.
\end{abstract}

\maketitle

\section{Introduction}\label{sec1}

Higher derivative theories has attracted physical interests from several different perspectives. In quantum gravity and cosmology, the higher derivative gravitational theories appear to be renormalizable~\cite{Stelle:1976gc} and provide feasible interpretations to the cosmological inflation~\cite{Starobinsky:1980te} and acceleration~\cite{Sotiriou:2008rp,Modesto:2011kw,Biswas:2011ar,Koshelev:2016xqb}. In particle physics, the Lee-Wick standard model provides a possible mechanics to stabilize quantum corrections on the Higgs mass~\cite{Lee:1969fy,Lee:1970iw,Grinstein:2007mp}.

The higher derivative theories can be dealt with Ostrogradsky's method~\cite{Ostrogradsky:1850fid,Woodard:2006nt} or Dirac's constraint method~\cite{Dirac:1950pj,Dirac:1951zz,Mannheim:2004qz}. For non-degenerated theories, these methods generally produce Hamiltonians being linearly dependent on some canonical momentum, which are recognized as the Ostrogradsky's instability. The quantum mechanics of higher derivative theories yields negative energy states or negative norm states, which imply the unbounded energy or the breaking of unitarity~\cite{Pais1950,Hawking:2001yt}. Kinds of approaches have been proposed to cure these problems~\cite{Woodard:2006nt,Salvio:2019ewf,Pavsic:2013noa,Pavsic:2016ykq}. In~\cite{Bender:2007wu}, non-Hermitian but $PT$ symmetrical Hamiltonian was introduced. On the other hand, alternative Hamiltonians~\cite{Masterov:2015ija} and super-symmetrical theories~\cite{Robert:2006nj} were considered. Besides, additional constraints were constructed to reduce the numbers of canonical variables in~\cite{Chen:2012au,Eliezer:1989cr}, in order to remove the unstable canonical momentums.

Another approach which can be used to analyze higher derivative theories is Caratheodory's equivalent Lagrangian method~\cite{Caratheodory}. Different from the methods of Ostrogradsky and Dirac, Caratheodory's method does not depend on the canonical variable and Hamiltonian. Instead, the Hamilton-Jacobi equation plays a critical role in Caratheodory's method. In Refs.~\cite{Bertin:2007gj,Pimentel:1995if}, higher derivative field theories have been treated with the Hamilton-Jacobi formalism derived through Caratheodory's method, which results are shown to be consistent with that derived from Dirac's formalism. 

Because of the peculiar feature that we can derive Hamilton-Jacobi equation without using canonical variables through Caratheodory's method, we shall show that there exist novel formulations of Hamilton-Jacobi equations for higher derivative theories. This novel formalism is able to degenerate into the conventional Hamilton-Jacobi equation for the first order theory.

As is well known, the Schr\"{o}dinger equation is closely related to the Hamilton-Jacobi equation. In the classical limit $\hbar\rightarrow 0$, the real part of Schr\"{o}dinger equation reproduces the Hamilton-Jacobi equation. On the contrary, once we have derived a novel formalism of Hamilton-Jacobi equation, we could attempt to seek the corresponding Schr\"{o}dinger equation which is able to yield the Hamilton-Jacobi equation in the classical limit. We shall show that the quantum mechanical correspondence of this novel Hamilton-Jacobi equation is nonlinear. The unbounded negative energy problem of higher derivative theories seems to be bypassed in this nonlinear  formalism.

In section~\ref{sec2}, as the preliminary setup and for the purpose of comparative investigations, we firstly discuss Ostrogradsky's method, which is employed to deal with the one dimensional mechanics system of higher order Lagrangian. We shall discuss the Hamilton-Jacobi equation in subsection~\ref{sec2a}, the Schr\"{o}dinger equation in subsection~\ref{sec2b} and their solutions. In section~\ref{sec3}, we shall derive the novel formalism of Hamilton-Jacobi equations for higher derivative theories using Caratheodory's method. Furthermore, we provide the nonlinear quantum mechanical correspondence for the novel Hamilton-Jacobi equation in subsection~\ref{sec3b} and discuss their solutions. In subsection~\ref{sec3bb}, we show that the unbounded negative energy problem could be resolved by considering the examples of the free particle potential and the harmonic oscillator potential. In section~\ref{sec4}, we shall discuss the higher dimensional mechanical system. Finally we present more discussions and conclusions in section~\ref{sec5}.

\section{Ostrogradsky's approach}\label{sec2}

In this section, we employ Ostrogradsky's method to deal with higher derivative mechanical systems in one dimensional space. The purpose of these discussions is to contrast with the future analysis of Caratheodory's method in section~\ref{sec3}. We consider the Lagrangian
\begin{eqnarray}
\label{sec2-osc}
L=-\frac{m\epsilon}{2\omega^2}\ddot{x}^2+\frac{1}{2}m\dot{x}^2-V(x),
\end{eqnarray}
in which the first term is a higher derivative term of second order~\cite{Woodard:2006nt}. We make the definitions $\dot{x}=\frac{dx}{dt}$ and 
$\ddot{x}=\frac{d^2x}{dt^2}$. $m$ is the mass parameter, and $\epsilon$ is the higher derivative coupling. $V(x)$  is the potential function. The equation of motion~(EOM) is 
\begin{eqnarray}
\label{sec2-osc-el}
\frac{m\epsilon}{\omega^2}x^{(4)}+m\ddot{x}+\frac{dV}{dx}=0.
\end{eqnarray}
In the below, we are going to use Ostrogradsky's method to derive the Hamiltonian for the Lagrangian~(\ref{sec2-osc}). We designate the new coordinate in configuration space
\begin{eqnarray}
\label{sec2-osc-ham-def}
y=\dot{x}.
\end{eqnarray}
With Ostrogradsky's method, the conjugate momentums for $x$ and $y$ are respectively
\begin{eqnarray}
\label{sec2-osc-ham-mom-x}
P_{x}&=&\frac{\partial L}{\partial \dot{x}}-\frac{d}{dt}\frac{\partial L}{\partial \ddot{x}}=m\dot{x}+\frac{m\epsilon}{\omega^2}\dddot{x},\\
\label{sec2-osc-ham-mom-y}
P_{y}&=&\frac{\partial L}{\partial \ddot{x}}=-\frac{m\epsilon}{\omega^2}\ddot{x}.
\end{eqnarray}
The Hamiltonian is computed as
\begin{eqnarray}
\label{sec2-osc-ham}
H=\dot{x}P_{x}+\dot{y}P_{y}-L=yP_{x}-\frac{m\epsilon}{2\omega^2}\ddot{x}^2-\frac{1}{2}my^2+V(x).
\end{eqnarray}
Represented with canonical variables, the Hamiltonian is
\begin{eqnarray}
\label{sec2-osc-ham-secon}
H=yP_{x}-\frac{\omega^2}{2m\epsilon}P^2_{y}-\frac{1}{2}my^2+V(x).
\end{eqnarray}
This Hamiltonian is linearly dependent on the momentum $P_x$, which is the incentive of Ostrogradsky's instability~\cite{Woodard:2006nt}. $(x,P_x)$ and $(y,P_y)$ are two pairs of canonical variables. The canonical equations of motion are
\begin{eqnarray}
\label{sec2-osc-ham-con-x}
\dot{x}&=&\frac{\partial H}{\partial P_{x}}=y,~~\dot{P}_{x}=-\frac{\partial H}{\partial {x}}=-\frac{dV}{dx},\\
\label{sec2-osc-ham-con-y}
\dot{y}&=&\frac{\partial H}{\partial P_{y}}=-\frac{\omega^2}{m\epsilon}P_{y},~~\dot{P}_{y}=-\frac{\partial H}{\partial {y}}=-P_{x}+m y.
\end{eqnarray}
The first equation is just the definition of the variable $y$. Introducing the action function $S(x,y,t)$, we obtain the Hamilton-Jacobi equation
\begin{eqnarray}
\label{sec2-osc-ham-jac}
\frac{\partial S}{\partial t}+H\left(x,y,P_x,P_y\right)=0.
\end{eqnarray}
together with the designations
\begin{eqnarray}
\label{sec2-osc-ham-jac-p}
P_x=\frac{\partial S}{\partial x}, ~~P_y=\frac{\partial S}{\partial y}.
\end{eqnarray}
Eq.~(\ref{sec2-osc-ham-jac}) is explicitly given as
\begin{eqnarray}
\label{sec2-osc-ham-jac-form}
\frac{\partial S}{\partial t}+y\frac{\partial S}{\partial x}-\frac{\omega^2}{2m\epsilon}\left(\frac{\partial S}{\partial y}\right)^2-\frac{1}{2}my^2+V(x)=0.
\end{eqnarray}
For the qualitative analysis of Eq.~(\ref{sec2-osc-ham-jac-form}), we will consider two specific examples in the following two subsections.

\subsection{Classical Solution of Hamilton-Jacobi Equation}\label{sec2a}

\subsubsection{Free Particle Potential}\label{sec2aa}

The first example is the free particle potential in one dimensional space~(1D). The potential function $V(x)$ is
\begin{eqnarray}
\label{sec2aa-pot}
V(x)=0.
\end{eqnarray}
In this case, the equation of motion~(\ref{sec2-osc-el}) is solved as
\begin{eqnarray}
\label{sec2aa-sol-x}
x(t)=a_{+}\cos(\tfrac{\omega}{\sqrt{\epsilon}}t)+a_{-}\sin(\tfrac{\omega}{\sqrt{\epsilon}}t)
+b_{+}t+b_{-},
\end{eqnarray}
where $a_{\pm}$ and $b_{\pm}$ are integral constants. This solution describes the motion of the free particle and the harmonic oscillator with the frequency $\tfrac{\omega}{\sqrt{\epsilon}}$ as two special cases respectively. The oscillation of frequency $\tfrac{\omega}{\sqrt{\epsilon}}$ is contributed by the higher derivative effect. 

The Hamilton-Jacobi equation~(\ref{sec2-osc-ham-jac-form}) has the variable-separating solution
\begin{eqnarray}
\label{sec2aa-sol-s}
S(t,x,y)&=&c_2-Et+m c_{1}x\\
&+&\frac{m}{2}\frac{\sqrt{\epsilon}}{\omega}\left\{\left(y-c_1\right)\sqrt{A^2-\left(y-c_1\right)^2}+A^2\arcsin\tfrac{1}{A}\left(y-c_1\right)\right\},\nonumber
\end{eqnarray}
where $c_{1}$, $c_{2}$ and $E$ are integral constants. $A$ is defined as
\begin{eqnarray}
\label{sec2aa-sol-a}
A=\sqrt{c_1^2-\tfrac{2E}{m}}.
\end{eqnarray}

\subsubsection{Harmonic Oscillator Potential}\label{sec2ab}

The second example is the harmonic oscillator in 1D. The potential function $V(x)$ is
\begin{eqnarray}
\label{sec2ab-pot}
V(x)=\frac{1}{2}m\omega^2x^2,
\end{eqnarray}
where $\omega$ is the constant frequency. The equation of motion~(EOM)~(\ref{sec2-osc-el}) has the general solution
\begin{eqnarray}
\label{sec2-osc-el-sol}
x(t)=a_{+}\sin(\omega_{+}t)+b_{+}\cos(\omega_{+}t)+a_{-}\sin(\omega_{-}t)+b_{-}\cos(\omega_{-}t),
\end{eqnarray}
where $\omega_{\pm}$ are defined as
\begin{eqnarray}
\label{sec2-osc-el-sol-ome}
\omega_{\pm}=\omega\sqrt{\frac{1\pm\sqrt{1-4\epsilon}}{2\epsilon}}.
\end{eqnarray}
This solution describes two harmonic oscillators with two different frequencies $\omega_{+}$ and $\omega_{-}$. when $\epsilon\rightarrow 0$, these two frequencies behave as
\begin{eqnarray}
\label{sec2-osc-el-sol-ome-ap}
\omega_{-}\rightarrow\omega,~~\omega_{+}\rightarrow\frac{\omega}{\sqrt{\epsilon}}.
\end{eqnarray}
For $\epsilon=\frac{1}{4}$, the frequencies $\omega_{+}$ and $\omega_{-}$ are degenerated, then we have
\begin{eqnarray}
\label{sec2-osc-el-ome}
\omega_{+}=\omega_{-}=\sqrt{2}\omega.
\end{eqnarray}
In this degenerated case, Eq.~(\ref{sec2-osc-el}) is solved as
\begin{eqnarray}
\label{sec2-osc-el-sol-fou}
x(t)=(a_{+}+a_{-}t)\cos(\sqrt{2}\omega t)+(b_{+}+b_{-}t)\sin(\sqrt{2}\omega t).
\end{eqnarray}

It is difficult to find the variable separating solution for the Hamilton-Jacobi equation~(\ref{sec2-osc-ham-jac-form}). Instead, Eq.~(\ref{sec2-osc-ham-jac-form}) can be solved by the polynomial ansatz
\begin{eqnarray}
\label{sec2a-osc-hj-form}
S(t,x,y)=\alpha(t)x^2+\beta(t)y^2+\chi(t)xy+\kappa(t)x+\sigma(t)y+\eta(t).
\end{eqnarray}
After inserting this expression into Eq.~(\ref{sec2-osc-ham-jac-form}), the coefficients of quadratic terms yield the equations
\begin{eqnarray}
\label{sec2a-osc-hj-alp}
\dot{\beta}+\chi-\frac{2\omega^2}{m\epsilon}\beta^2-\frac{m}{2}&=&0,\\
\label{sec2a-osc-hj-bet}
\dot{\chi}+2\alpha-\frac{2\omega^2}{m\epsilon}\chi\beta&=&0,\\
\label{sec2a-osc-hj-lam}
\dot{\alpha}-\frac{\omega^2}{2m\epsilon}\chi^2+\frac{1}{2}m\omega^2&=&0,
\end{eqnarray}
where $\dot{\alpha}=\frac{d\alpha}{dt}$, $\dot{\beta}=\frac{d\beta}{dt}$ and so forth.
The coefficients of linear terms and the constant term yield the equations
\begin{eqnarray}
\label{sec2a-osc-hj-rho}
\dot{\eta}-\frac{\omega^2}{2m\epsilon}\sigma^2&=&0,\\
\label{sec2a-osc-hj-sig}
\dot{\sigma}+\kappa-\frac{2\omega^2}{m\epsilon}\beta\sigma&=&0,\\
\label{sec2a-osc-hj-con}
\dot{\kappa}-\frac{\omega^2}{m\epsilon}\chi\sigma&=&0.
\end{eqnarray}
With Eq.~(\ref{sec2a-osc-hj-alp}), $\chi(t)$ can be expressed with $\beta(t)$
\begin{eqnarray}
\label{sec2a-osc-hj-alp-b}
\chi=-\dot{\beta}+\frac{2\omega^2}{m\epsilon}\beta^2+\frac{m}{2}.
\end{eqnarray}
With Eqs.~(\ref{sec2a-osc-hj-alp-b}) and (\ref{sec2a-osc-hj-bet}),  $\alpha(t)$ is solved as
\begin{eqnarray}
\label{sec2a-osc-hj-bet-b}
\alpha=\frac{1}{2}\ddot{\beta}-\frac{3\omega^2}{m\epsilon}\beta\dot{\beta}+\frac{2\omega^4}{m^2\epsilon^2}\beta^3+\frac{\omega^2}{2\epsilon}\beta.
\end{eqnarray}
Finally, inserting $\alpha(t)$ and $\chi(t)$ into Eq.~(\ref{sec2a-osc-hj-lam}), we obtain an single equation of $\beta(t)$
\begin{eqnarray}
\label{sec2a-osc-hj-lam-b}
\frac{1}{2}\dddot{\beta}-\frac{3\omega^2}{m\epsilon}(\beta\dot{\beta}+\dot{\beta}^2)+\frac{6\omega^4}{m^2\epsilon^2}\beta^2\dot{\beta}+\frac{\omega^2}{2\epsilon}\dot{\beta}-\frac{\omega^2}{2m\epsilon}\left(-\dot{\beta}+\frac{2\omega^2}{m\epsilon}\beta^2+\frac{m}{2}\right)^2+\frac{1}{2}m\omega^2=0.
\end{eqnarray}
Using the transformation
\begin{eqnarray}
\label{sec2a-osc-hj-beta-tr}
\beta=-\frac{m\epsilon}{2\omega^2}\frac{\dot{\vartheta}}{\vartheta},
\end{eqnarray}
Eq.~(\ref{sec2a-osc-hj-lam-b}) becomes
\begin{eqnarray}
\label{sec2a-osc-hj-theta}
\frac{\epsilon}{\omega^2}\left(\frac{\vartheta^{(4)}}{\vartheta}-\frac{\dot{\vartheta}\vartheta^{(3)}}{\vartheta^2}+\frac{1}{2}\frac{\ddot{\vartheta}^2}{\vartheta^2}\right)+2\frac{\ddot{\vartheta}}{\vartheta}-\frac{\dot{\vartheta}^2}{\vartheta^2}+\frac{1-4\epsilon}{2\epsilon}\omega^2=0.
\end{eqnarray}
This equation can be solved as
\begin{eqnarray}
\label{sec2a-osc-hj-theta-sol}
\vartheta(t)=c_{+}+a_{+}\cos({\nu_{+}}t)+b_{+}\sin({\nu_{+}}t)+a_{-}\cos({\nu_{-}}t)+b_{-}\sin({\nu_{-}}t),
\end{eqnarray}
where
\begin{eqnarray}
\label{sec2a-osc-hj-theta-sol-nu}
\nu_{\pm}&=&\omega\sqrt{\frac{1\pm2\sqrt{\epsilon}}{\epsilon}},\\
\label{sec2a-osc-hj-theta-sol-c}
c_{+}&=&2\omega\sqrt{\tfrac{a_{-}^2+b_{-}^2}{\nu^2_{+}\sqrt{\epsilon}}-\tfrac{a_{+}^2+b_{+}^2}{\nu^2_{-}\sqrt{\epsilon}}}.
\end{eqnarray}
We can verify the following identities
\begin{eqnarray}
\label{sec2a-osc-hj-theta-sol-nu-ome}
\nu_{+}=\omega_{+}+\omega_{-},~\nu_{-}=\omega_{+}-\omega_{-}.
\end{eqnarray}
Using the solution of $\vartheta(t)$ in Eq.~(\ref{sec2a-osc-hj-theta-sol}), we can work out the solutions for other unknown functions. Then the solution of Eq.~(\ref{sec2a-osc-hj-form}) can be expressed by the function $\vartheta(t)$.

\subsection{Quantum mechanics}\label{sec2b}

$(P_{x},x)$ and $(P_{y},y)$ are two pairs of canonical variables. The quantum mechanical equation of Eq.~(\ref{sec2-osc-ham-jac-form}) can be derived through the canonical quantization. Using the coordinate representations
\begin{eqnarray}
\label{sec2-osc-sch-rep}
P_{x}\rightarrow -i\hbar\frac{\partial}{\partial x},~~P_{y}\rightarrow -i\hbar\frac{\partial}{\partial y},
\end{eqnarray}
we can obtain the Schr\"{o}dinger equation
\begin{eqnarray}
\label{sec2-osc-sch-or}
i\hbar\frac{\partial \Psi}{\partial t}=-i\hbar y\frac{\partial \Psi}{\partial x}+\frac{\hbar^2\omega^2}{2m\epsilon}\frac{\partial^2 \Psi}{\partial y^2}
-\frac{1}{2}my^2\Psi+V(x)\Psi.
\end{eqnarray}
Divided by $\Psi$, Eq.~(\ref{sec2-osc-sch-or}) can be converted into
\begin{eqnarray}
\label{sec2-osc-sch}
i\hbar\frac{1}{\Psi} \frac{\partial \Psi}{\partial t}=-i\hbar y\frac{1}{\Psi} \frac{\partial \Psi}{\partial x}+\frac{\hbar^2\omega^2}{2m\epsilon}\frac{1}{\Psi} \frac{\partial^2 \Psi}{\partial y^2}
-\frac{1}{2}my^2+V(x).
\end{eqnarray}
The real part of Eq.~(\ref{sec2-osc-sch}) is
\begin{eqnarray}
\label{sec2-osc-sch-real}
-\hbar\mathbf{Im}\left(\tfrac{1}{\Psi} \tfrac{\partial \Psi}{\partial t}\right)=\hbar y\mathbf{Im}\left(\tfrac{1}{\Psi} \tfrac{\partial \Psi}{\partial x}\right)
+\frac{\hbar^2\omega^2}{2m\epsilon}\mathbf{Re}\left(\tfrac{1}{\Psi} \tfrac{\partial^2 \Psi}{\partial y^2}\right)-\frac{1}{2}my^2+V(x),
\end{eqnarray}
and the imaginary part of Eq.~(\ref{sec2-osc-sch}) is
\begin{eqnarray}
\label{sec2-osc-sch-ima}
\hbar\mathbf{Re}\left(\tfrac{1}{\Psi} \tfrac{\partial \Psi}{\partial t}\right)=-\hbar y\mathbf{Re}\left(\tfrac{1}{\Psi}\tfrac{\partial \Psi}{\partial x}\right)
+\tfrac{\hbar^2\omega^2}{2m\epsilon}\mathbf{Im}\left(\tfrac{1}{\Psi}\tfrac{\partial^2 \Psi}{\partial y^2}\right).
\end{eqnarray}
In the above, the real part and imaginary part are defined as
\begin{eqnarray}
\label{sec2-osc-v-re}
\mathbf{Im}\left(\tfrac{1}{\Psi}\tfrac{\partial\Psi}{\partial t}\right)=\tfrac{1}{2i}\left(\tfrac{1}{\Psi}\tfrac{\partial\Psi}{\partial t}-\tfrac{1}{\Psi^{*}}\tfrac{\partial\Psi^{*}}{\partial t}\right),~\mathbf{Re}\left(\tfrac{1}{\Psi}\tfrac{\partial^2\Psi}{\partial y^2}\right)
=\tfrac{1}{2}\left(\tfrac{1}{\Psi}\tfrac{\partial^2\Psi}{\partial y^2}+\tfrac{1}{\Psi^{*}}\tfrac{\partial^2\Psi^{*}}{\partial y^2}\right),
\end{eqnarray}
where $\Psi^{*}$ stands for the complex conjugate of $\Psi$. $\Psi$ can be decomposed as
\begin{eqnarray}
\label{sec2-osc-psi}
\Psi=R(x,y,t) e^{\frac{i}{\hbar}S(x,y,t)},
\end{eqnarray}
where $R$ and $S$ are both real functions. We have the expansions
\begin{eqnarray}
\label{sec2-osc-psi-t}
i\hbar\frac{1}{\Psi} \frac{\partial \Psi}{\partial t}&=&-\frac{\partial S}{\partial t}+i\hbar\frac{1}{R} \frac{\partial R}{\partial t},\\
\label{sec2-osc-psi-yy}
-\hbar^2\frac{1}{\Psi} \frac{\partial^2 \Psi}{\partial y^2}&=&\left(\frac{\partial S}{\partial y}\right)^2-\hbar^2\frac{1}{R} \frac{\partial^2 R}{\partial y^2}-i\hbar\frac{\partial^2 S}{\partial y^2}-2i\hbar\frac{1}{R} \frac{\partial R}{\partial y}\frac{\partial S}{\partial y}.
\end{eqnarray}
With these two expressions, Eqs.~(\ref{sec2-osc-sch-real}) and (\ref{sec2-osc-sch-ima}) are respectively 
\begin{eqnarray}
\label{sec2-osc-sch-real-a}
\frac{\partial S}{\partial t}+y\frac{\partial S}{\partial x}-\frac{\omega^2}{2m\epsilon}\left(\frac{\partial S}{\partial y}\right)^2-\frac{1}{2}my^2+V(x)+\frac{\hbar^2\omega^2}{2m\epsilon}\frac{1}{R} \frac{\partial^2 R}{\partial y^2}&=&0,\\
\label{sec2-osc-sch-ima-a}
\frac{1}{R} \frac{\partial R}{\partial t}+y\frac{1}{R} \frac{\partial R}{\partial x}-\frac{\omega^2}{2m\epsilon }\frac{\partial^2 S}{\partial y^2}-\frac{\omega^2}{m\epsilon}\frac{1}{R} \frac{\partial R}{\partial y}\frac{\partial S}{\partial y}&=&0.
\end{eqnarray}
With these two equations, we can find the connections between classical mechanics and quantum mechanics. Eq.~(\ref{sec2-osc-sch-real-a}) is the Hamilton-Jacobi equation (\ref{sec2-osc-ham-jac-form}) plus Bohm's quantum potential~\cite{Bohm:1951xw}
\begin{eqnarray}
\label{sec2-osc-sch-bohm}
U_{q}=\frac{\hbar^2\omega^2}{2m\epsilon}\frac{1}{R} \frac{\partial^2 R}{\partial y^2}.
\end{eqnarray}
In the classical limit $\hbar\rightarrow 0$, the quantum potential vanishes, then Eq.~(\ref{sec2-osc-sch-real-a}) reduces to the Hamilton-Jacobi equation (\ref{sec2-osc-ham-jac-form}). Eq.~(\ref{sec2-osc-sch-ima-a}) can be rewritten as
\begin{eqnarray}
\label{sec2-osc-sch-ima-a-b}
\frac{\partial \rho}{\partial t}+\frac{\partial }{\partial x}(\rho v_{x})+\frac{\partial }{\partial y}(\rho v_{y})=0,
\end{eqnarray}
where
\begin{eqnarray}
\label{sec2-osc-sch-ima-a-b-def}
\rho(x,y,t)=R^2=\Psi^{*}\Psi,~v_{x}=y,~v_{y}=-\frac{\omega^2}{m\epsilon}\frac{\partial S}{\partial y}.
\end{eqnarray}
$v_x$ and $v_y$ are the velocity components of probability density fluid. Eq.~(\ref{sec2-osc-sch-ima-a-b}) is the continuity equation, which implies the conservation of probability. 

Now we begin to discuss the solutions of Eq.~(\ref{sec2-osc-sch}). We consider the stationary solution
\begin{eqnarray}
\label{sec2-osc-sch-sta}
\Psi(t,x,y)=\psi(x,y)e^{-\frac{i}{\hbar}Et}.
\end{eqnarray}
With this ansatz, Eq.~(\ref{sec2-osc-sch}) becomes
\begin{eqnarray}
\label{sec2-osc-sch-sta-for}
E-V(x)+\frac{1}{2}my^2=-i\hbar y\frac{1}{\psi} \frac{\partial \psi}{\partial x}+\frac{\hbar^2\omega^2}{2m\epsilon}\frac{1}{\psi} \frac{\partial^2 \psi}{\partial y^2}.
\end{eqnarray}

\subsubsection{Free Particle Potential}\label{sec2ba}

For the free particle potential~(\ref{sec2aa-pot}), we obtain from Eq.~(\ref{sec2-osc-sch-sta-for})
\begin{eqnarray}
\label{sec2ba-sch}
E+\frac{1}{2}my^2=-i\hbar y\frac{1}{\psi} \frac{\partial \psi}{\partial x}+\frac{\hbar^2\omega^2}{2m\epsilon}\frac{1}{\psi} \frac{\partial^2 \psi}{\partial y^2}, 
\end{eqnarray}
which can be solved by the method of separating variables
\begin{eqnarray}
\label{sec2ba-sch-ant}
\psi(x,y)=\eta(x)\varphi(y).
\end{eqnarray}
For $\eta(x)$, we have the equation
\begin{eqnarray}
\label{sec2ba-sch-eq-t}
\frac{1}{\eta}\frac{d\eta}{dx}=\frac{i}{\hbar}my_0,
\end{eqnarray}
where $y_0$ is a constant number. For $\varphi(y)$, we obtain the equation
\begin{eqnarray}
\label{sec2ba-sch-eq-p}
\frac{\hbar^2\omega^2}{2m\epsilon}\frac{d^2\varphi}{dy^2}-\left[\frac{1}{2}my^2-my_0y+E\right]\varphi=0.
\end{eqnarray}
The solution of $\eta(x)$ is
\begin{eqnarray}
\label{sec2ba-sch-t-s}
\eta(x)=c_3 e^{\tfrac{i}{\hbar}my_0x},
\end{eqnarray}
which is the plane wave solution, and $c_3$ is the integral constant. The solution for $\varphi(y)$ is
\begin{eqnarray}
\label{sec2ba-sch-ph-s}
\varphi(y)=c_1 e^{-\tfrac{z^2}{2}}\mathrm{hypergeom}\left(\tfrac{1-\lambda}{4},\tfrac{1}{2},z^2\right)+c_2 z e^{-\tfrac{z^2}{2}}\mathrm{hypergeom}\left(\tfrac{3-\lambda}{4},\tfrac{3}{2},z^2\right),
\end{eqnarray}
where $c_1$ and $c_2$ are integral constants. We have defined
\begin{eqnarray}
\label{sec2ba-sch-ph-s-def}
z=\sqrt{\tfrac{m\sqrt{\epsilon}}{\hbar\omega}}(y-y_0),~~
\lambda=\left(-\tfrac{2E}{m}+y_0^2\right)\tfrac{m\sqrt{\epsilon}}{\hbar\omega}.
\end{eqnarray}
For the hypergeometrical function being convergent at infinity, we need the relation
\begin{eqnarray}
\label{sec2ba-sch-ph-s-lam}
\lambda=2n+1, ~~n=0,1,2,\cdots.
\end{eqnarray}
Together with Eq.~(\ref{sec2ba-sch-ph-s-def}), this constraint yields 
\begin{eqnarray}
\label{sec2ba-sch-ph-s-e}
E=\frac{1}{2}my_0^2-(n+\frac{1}{2})\hbar\frac{\omega}{\sqrt{\epsilon}}.
\end{eqnarray}
In this expression of the eigenenergy $E$, the part $\frac{1}{2}my_0^2$ can be interpreted as the kinetic energy of the free particle. $(n+\frac{1}{2})\hbar\frac{\omega}{\sqrt{\epsilon}}$ corresponds to the quantum eigenenergy of the harmonic oscillator with the frequency $\frac{\omega}{\sqrt{\epsilon}}$. From the above, we have seen that the harmonic oscillator gives negative contribution to the total energy, which causes the total energy unbounded from below. This is known as the Ostrogradsky's instability, which is induced by the higher derivative effect associated with $\epsilon$.

\subsubsection{Harmonic Oscillator Potential}\label{sec2bb}

For the harmonic potential (\ref{sec2ab-pot}), we obtain from Eq.~(\ref{sec2-osc-sch-sta-for})
\begin{eqnarray}
\label{sec2bb-sch}
E-\frac{1}{2}m\omega^2x^2+\frac{1}{2}my^2=-i\hbar y\frac{1}{\psi} \frac{\partial \psi}{\partial x}+\frac{\hbar^2\omega^2}{2m\epsilon}\frac{1}{\psi} \frac{\partial^2 \psi}{\partial y^2}, 
\end{eqnarray}
At first, we consider several special solutions of Eq.~(\ref{sec2bb-sch}). One typical solution of Eq.~(\ref{sec2bb-sch}) is given as
\begin{eqnarray}
\label{sec2bb-sol}
E=\frac{1}{2}\hbar\nu_{-}=\frac{1}{2}\hbar(\omega_{+}-\omega_{-}),~\psi(x,y)=e^{i\frac{m\sqrt{\epsilon}}{\hbar}xy}\exp\left\{\frac{m}{2\hbar}\nu_{-}\sqrt{\epsilon}\left(x^2+\frac{\sqrt{\epsilon}}{\omega^2}y^2\right)\right\},
\end{eqnarray}
where $\nu_{-}$ is defined in Eq.~(\ref{sec2a-osc-hj-theta-sol-nu}), or
\begin{eqnarray}
\label{sec2bb-sol-b}
E=-\frac{1}{2}\hbar\nu_{-}=-\frac{1}{2}\hbar(\omega_{+}-\omega_{-}),~\psi(x,y)=e^{i\frac{m\sqrt{\epsilon}}{\hbar}xy}
\exp\left\{-\frac{m}{2\hbar}\nu_{-}\sqrt{\epsilon}\left(x^2+\frac{\sqrt{\epsilon}}{\omega^2}y^2\right)\right\}.
\end{eqnarray}
Another typical solution of Eq.~(\ref{sec2bb-sch}) is
\begin{eqnarray}
\label{sec2bb-sol-2a}
E=\frac{1}{2}\hbar\nu_{+}=\frac{1}{2}\hbar(\omega_{+}+\omega_{-}),~\psi(x,y)=e^{-i\frac{m\sqrt{\epsilon}}{\hbar}xy}\exp\left\{-\frac{m}{2\hbar}\nu_{+}\sqrt{\epsilon}\left(x^2-\frac{\sqrt{\epsilon}}{\omega^2}y^2\right)\right\},
\end{eqnarray}
where $\nu_{+}$ is defined in Eq.~(\ref{sec2a-osc-hj-theta-sol-nu}), or
\begin{eqnarray}
\label{sec2bb-sol-2b}
E=-\frac{1}{2}\hbar\nu_{+}=-\frac{1}{2}\hbar(\omega_{+}+\omega_{-}),~\psi(x,y)=e^{-i\frac{m\sqrt{\epsilon}}{\hbar}xy}\exp\left\{\frac{m}{2\hbar}\nu_{+}\sqrt{\epsilon}\left(x^2-\frac{\sqrt{\epsilon}}{\omega^2}y^2\right)\right\}.
\end{eqnarray}
The solutions in Eqs.~(\ref{sec2bb-sol}), (\ref{sec2bb-sol-2a}) and (\ref{sec2bb-sol-2b}) are not convergent at infinity, which are not normalizable. 

In order to derive general solutions of Eq.~(\ref{sec2bb-sch}), we consider the transformation
\begin{eqnarray}
\label{sec2bb-tr}
\psi(x,y)=e^{i\frac{m\sqrt{\epsilon}}{\hbar}xy}
\exp\left\{-\frac{m}{2\hbar}\nu_{-}\sqrt{\epsilon}\left(x^2+\frac{\sqrt{\epsilon}}{\omega^2}y^2\right)\right\}\phi(z,\tau),
\end{eqnarray}
where
\begin{eqnarray}
\label{sec2bb-tr-z}
z=\omega_{-}x+iy,~~\tau=\omega_{+}x-iy.
\end{eqnarray}
Then from Eq.~(\ref{sec2bb-sch}), we obtain 
\begin{eqnarray}
\label{sec2bb-tr-re}
-\frac{\omega^2\hbar^2}{2m\epsilon}\left(\frac{\partial^2}{\partial z^2}-2\frac{\partial^2}{\partial z \partial \tau}+\frac{\partial^2}{\partial \tau^2}\right)\phi(z,\tau)&+&\hbar\left(\omega_{-}\tau\frac{\partial}{\partial \tau}-\omega_{+}z\frac{\partial}{\partial z}\right)\phi(z,\tau)\\
&=&E\phi(z,\tau)+\frac{\hbar}{2}(\omega_{+}-\omega_{-})\phi(z,\tau).\nonumber
\end{eqnarray}
The variables can be separated as
\begin{eqnarray}
\label{sec2bb-tr-z-s}
\phi(z,\tau)=\phi_{+}(z)+\phi_{-}(\tau).
\end{eqnarray}
The first typical solutions for $\phi_{+}(z)$ and $\phi_{-}(\tau)$ are
\begin{eqnarray}
\label{sec2bb-sch-sol-a}
\phi_{+}(z)&=&c_1 \mathrm{hypergeom}\left(\tfrac{\hbar\omega_{+}-\hbar\omega_{-}+2E}{4\hbar\omega_{+}},\tfrac{1}{2},-\tfrac{\omega_{+}}{\omega}\tfrac{m\epsilon}{\omega\hbar}z^2\right)\\
&+&c_2 z\cdot\mathrm{hypergeom}\left(\tfrac{3\hbar\omega_{+}-\hbar\omega_{-}+2E}{4\hbar\omega_{+}},\tfrac{3}{2},-\tfrac{\omega_{+}}{\omega}\tfrac{m\epsilon}{\omega\hbar}z^2\right),\nonumber\\
\label{sec2bb-sch-sol-a-b}
\phi_{-}(\tau)&=&0.
\end{eqnarray}
The second typical solutions are
\begin{eqnarray}
\label{sec2bb-sch-sol-b}
\phi_{-}(\tau)&=&c_3 \mathrm{hypergeom}\left(\tfrac{\hbar\omega_{-}-\hbar\omega_{+}-2E}{4\hbar\omega_{-}},\tfrac{1}{2},\tfrac{\omega_{-}}{\omega}\tfrac{m\epsilon}{\omega\hbar}\tau^2\right)\\
&+&c_4 \tau\cdot\mathrm{hypergeom}\left(\tfrac{3\hbar\omega_{-}-\hbar\omega_{+}-2E}{4\hbar\omega_{-}},\tfrac{3}{2},\tfrac{\omega_{-}}{\omega}\tfrac{m\epsilon}{\omega\hbar}\tau^2\right),\nonumber\\
\label{sec2bb-sch-sol-b}
\phi_{+}(z)&=&0.
\end{eqnarray}
In the above, $c_1$, $c_2$, $c_3$ and $c_4$ are integral constants. For $\phi_{+}(z)$ being finite when $z\rightarrow\infty$, we need the condition
\begin{eqnarray}
\label{sec2bb-sch-sol-a-c}
E_{+}=-n\hbar\omega_{+}+\tfrac{1}{2}\hbar(\omega_{-}-\omega_{+}),~~n=0,1,2,\cdots.
\end{eqnarray}
For $\phi_{-}(\tau)$ being finite when $\tau\rightarrow\infty$, we need the condition
\begin{eqnarray}
\label{sec2bb-sch-sol-b-c}
E_{-}=m\hbar\omega_{-}+\tfrac{1}{2}\hbar(\omega_{-}-\omega_{+}),~~m=0,1,2,\cdots.
\end{eqnarray}
The energy $E_{+}$ is unbounded from below. Moreover, Eq.~(\ref{sec2bb-tr-re}) admits more general solutions with the eigenenergy
\begin{eqnarray}
\label{sec2bb-sch-sol-b-c-m}
E=\left(m+\tfrac{1}{2}\right)\hbar\omega_{-}-\left(n+\tfrac{1}{2}\right)\hbar\omega_{+}.
\end{eqnarray}
The more general eigensolutions of Eq.~(\ref{sec2bb-tr-re}) in the Pais-Uhlenbeck formalism~\cite{Pais1950} can be found in~\cite{Smilga:2005gb,Pavsic:2013noa,Pavsic:2016ykq}.

\section{Caratheodory's approach}\label{sec3} 

In the forgoing section~\ref{sec2}, we have treated the higher derivative mechanical models with Ostrogradsky's method. On the other hand, these models can also been dealt with Dirac's constraint method as in~\cite{Mannheim:2004qz,Bender:2007wu}. These two different methods generally yield equivalent results. The common feature of these two methods is the employment of Hamilton's canonical variables. In this section, we shall use Caratheodory's method to analyze the higher derivative theories in one dimensional space.

The particular feature of Caratheodory's method is that we can derive the Hamilton-Jacobi equation using the Lagrangian, but without using the Hamiltonian.

For some Lagrangian $L$, we can perform variation calculus to find its equation of motion~(EOM). However, the EOM is not affected if we add some total divergence term to the Lagrangian. Caratheodory proposed that we can add the total divergence term to the Lagrangian such that
\begin{eqnarray}
\label{sec3-osc-cara}
L(x,\dot{x})-\frac{dS(x,t)}{dt}=0.
\end{eqnarray}
Here $\frac{dS}{dt}$ is the total divergence term. We expand Eq.~(\ref{sec3-osc-cara}) as
\begin{eqnarray}
\label{sec3-osc-cara-re}
L(x,\dot{x})=\frac{\partial S}{\partial t}+\dot{x}\frac{\partial S}{\partial x}.
\end{eqnarray}
From Eq.~(\ref{sec3-osc-cara-re}), we obtain
\begin{eqnarray}
\label{sec3-osc-cara-re-dx}
\frac{\partial L}{\partial \dot{x}}=\frac{\partial S}{\partial x},
\end{eqnarray}
which can be used to express $\dot{x}$ in terms of $\frac{\partial S}{\partial x}$ such as
\begin{eqnarray}
\label{sec3-osc-cara-re-dx-sol}
\dot{x}=v\left(x,\tfrac{\partial S}{\partial x}\right).
\end{eqnarray}
Inserting this expression of  $\dot{x}$ into Eq.~(\ref{sec3-osc-cara-re}), we can obtain the Hamilton-Jacobi equation
\begin{eqnarray}
\label{sec3-osc-cara-re-hj}
\frac{\partial S}{\partial t}+v\frac{\partial S}{\partial x}=L(x,v).
\end{eqnarray}
The foregoing transformation in Eq.~(\ref{sec3-osc-cara-re-dx-sol}) is similar to the Legendre transformation. 

The above discussions are implemented for the Lagrangian with first order derivatives. For the Lagrangian with second order derivatives, we can consider the following surface terms such that
\begin{eqnarray}
\label{sec3-osc-cara-2nd}
L(x,\dot{x},\ddot{x})-\frac{d F(x,t)}{dt}-\frac{d}{dt}\left(\dot{x}f(x,t)\right)=0,
\end{eqnarray}
where $F(x,t)$ and $f(x,t)$ are two undetermined functions. The surface terms in Eq.~(\ref{sec3-osc-cara-2nd}) are not the usual constructions. There are different constructions presented in~\cite{Pimentel:1995if}. Eq.~(\ref{sec3-osc-cara-2nd}) can be computed as
\begin{eqnarray}
\label{sec3-osc-cara-2nd-re}
L(x,\dot{x},\ddot{x})=\frac{\partial F}{\partial t}+\dot{x}\frac{\partial F}{\partial x}+\ddot{x}f+\dot{x}\left(\frac{\partial f}{\partial t}+\dot{x}\frac{\partial f}{\partial x}\right).
\end{eqnarray}
For the Lagrangian with third order derivatives, we can consider the formulation
\begin{eqnarray}
\label{sec3-osc-cara-3nd}
L(x,\dot{x},\ddot{x},\dddot{x})-\frac{d F(x,t)}{dt}-\frac{d}{dt}\left(\dot{x}f(x,t)\right)-\frac{d}{dt}\left(\ddot{x}h(x,t)\right)=0,
\end{eqnarray}
where we have introduced a new function $h(x,t)$. The aforementioned construction can be generalized to the Lagrangian with higher order derivatives.

From Eq.~(\ref{sec3-osc-cara-2nd-re}), we obtain
\begin{eqnarray}
\label{sec3-osc-cara-2nd-dx}
\frac{\partial L}{\partial \dot{x}}&=&\frac{\partial F}{\partial x}+\frac{\partial f}{\partial t}+2\dot{x}\frac{\partial f}{\partial x},\\
\label{sec3-osc-cara-2nd-2dx}
\frac{\partial L}{\partial \ddot{x}}&=&f.
\end{eqnarray}
From Eqs.~(\ref{sec3-osc-cara-2nd-dx}) and (\ref{sec3-osc-cara-2nd-2dx}), we can solve $\dot{x}$ and $\ddot{x}$ in terms of $\frac{\partial F}{\partial x}$, $\frac{\partial f}{\partial t}$, $\frac{\partial f}{\partial x}$ and $f$ such as
\begin{eqnarray}
\label{sec3-osc-cara-2nd-dx-sol}
\dot{x}&=&v\left(t,x,\tfrac{\partial F}{\partial x}, \tfrac{\partial f}{\partial t}, \tfrac{\partial f}{\partial x},f\right),\\
\label{sec3-osc-cara-2nd-2dx-sol}
\ddot{x}&=&a\left(t,x,\tfrac{\partial F}{\partial x}, \tfrac{\partial f}{\partial t}, \tfrac{\partial f}{\partial x},f\right).
\end{eqnarray}
Inserting these expressions of  $\dot{x}$ and $\ddot{x}$ into Eq.~(\ref{sec3-osc-cara-2nd-re}), we can obtain the Hamilton-Jacobi equation
\begin{eqnarray}
\label{sec3-osc-cara-2nd-hjz}
\frac{\partial F}{\partial t}+v\frac{\partial F}{\partial x}+af+v\cdot\left(\frac{\partial f}{\partial t}+v\frac{\partial f}{\partial x}\right)=L(x,v,a),
\end{eqnarray}
in which there are two undetermined functions $F(x,t)$ and $f(x,t)$. However, for theories of second order derivatives, the functions $v$ and $a$ are not independent. The definitions of $\dot{x}$ and $\ddot{x}$ provide the relation between $v$ and $a$
\begin{eqnarray}
\label{sec3-osc-cara-2nd-cons}
\ddot{x}=\frac{d}{dt}\dot{x}=\frac{d v(x,t)}{dt}=\frac{\partial v(x,t)}{\partial t}+\dot{x}\frac{\partial v(x,t)}{\partial x}=\frac{\partial v}{\partial t}+v\frac{\partial v}{\partial x},
\end{eqnarray}
which is
\begin{eqnarray}
\label{sec3-osc-cara-2nd-cons-a}
a(x,t)=\frac{\partial v}{\partial t}+v\frac{\partial v}{\partial x},
\end{eqnarray}
which is reminiscent of Euler's equation in fluid dynamics. Eq.~(\ref{sec3-osc-cara-2nd-cons-a}) establishes an additional constraint on $F(x,t)$ and $f(x,t)$. Using this constraint, we can solve $f(x,t)$ in terms of $F(x,t)$, so that there is only one undetermined function $F(x,t)$ left in the Hamilton-Jacobi equation~(\ref{sec3-osc-cara-2nd-hjz}). 

Eqs.~(\ref{sec3-osc-cara-2nd-hjz}) and (\ref{sec3-osc-cara-2nd-cons-a}) are the Hamilton-Jacobi equations derived through Caratheodory's method. We shall show that they have particularly novel formulations, which are different from the canonical Hamilton-Jacobi equations presented in section~\ref{sec2}.

As an example, we deal with the second-order Lagrangian~(\ref{sec2-osc}). Using Eq.~(\ref{sec3-osc-cara-2nd-re}), we have
\begin{eqnarray}
\label{sec3-osc}
-\frac{m\epsilon}{2\omega^2}\ddot{x}^2+\frac{1}{2}m\dot{x}^2-V(x)=\frac{\partial F}{\partial t}+\dot{x}\frac{\partial F}{\partial x}
+\ddot{x}f+\dot{x}\left(\frac{\partial f}{\partial t}+\dot{x}\frac{\partial f}{\partial x}\right).
\end{eqnarray}
From Eq.~(\ref{sec3-osc-cara-2nd-dx}) and (\ref{sec3-osc-cara-2nd-2dx}), we obtain
\begin{eqnarray}
\label{sec3-cara-2nd-dx-eq}
m\dot{x}&=&\frac{\partial F}{\partial x}+\frac{\partial f}{\partial t}+2\dot{x}\frac{\partial f}{\partial x},\\
\label{sec3-cara-2nd-2dx-eq}
-\frac{m\epsilon}{\omega^2}\ddot{x}&=&f.
\end{eqnarray}
From these two equations, $\dot{x}$ and $\ddot{x}$ can be solved as
\begin{eqnarray}
\label{sec3-cara-2nd-dx-sol}
\dot{x}&=&v(x,t)=\frac{1}{m-2\frac{\partial f}{\partial x}}\left(\frac{\partial F}{\partial x}+\frac{\partial f}{\partial t}\right),\\
\label{sec3-cara-2nd-2dx-sol}
\ddot{x}&=&a(x,t)=-\frac{\omega^2}{m\epsilon}f.
\end{eqnarray}
If we insert the expressions of Eq.~(\ref{sec3-cara-2nd-dx-sol}) and (\ref{sec3-cara-2nd-2dx-sol}) into the Hamilton-Jacobi equation (\ref{sec3-osc}) and the constraint equation (\ref{sec3-osc-cara-2nd-cons-a}), we will obtain complicated and very lengthy formulations. On the other hand, there is an approach to simplify the aforementioned results.

Using the definition of $\dot{x}$ in Eq.~(\ref{sec3-osc-cara-2nd-dx-sol}) and the definition of 
$\ddot{x}$ in Eq.~(\ref{sec3-osc-cara-2nd-cons}), we obtain
\begin{eqnarray}
\label{sec3-osc-cara-2nd-dx-sol-a}
\dot{x}&=&v\left(x,t\right),\\
\label{sec3-osc-cara-2nd-cons-aa}
\ddot{x}&=&\frac{\partial v}{\partial t}+v\frac{\partial v}{\partial x}.
\end{eqnarray}
Then Eqs.~(\ref{sec3-cara-2nd-dx-eq}) and (\ref{sec3-cara-2nd-2dx-eq}) turn into
\begin{eqnarray}
\label{sec3-cara-2nd-dx-sol-re}
mv&=&\frac{\partial F}{\partial x}+\frac{\partial f}{\partial t}+2v\frac{\partial f}{\partial x},\\
\label{sec3-cara-2nd-2dx-sol-re}
f&=&-\frac{m\epsilon}{\omega^2}\left(\frac{\partial v}{\partial t}+v\frac{\partial v}{\partial x}\right).
\end{eqnarray}
Eq.~(\ref{sec3-osc}) becomes
\begin{eqnarray}
\label{sec3-osc-re}
-\frac{m\epsilon}{2\omega^2}\left(\frac{\partial v}{\partial t}+v\frac{\partial v}{\partial x}\right)^2&+&\frac{1}{2}mv^2-V(x)=\frac{\partial F}{\partial t}+v\frac{\partial F}{\partial x}\\
&+&\left(\frac{\partial v}{\partial t}+v\frac{\partial v}{\partial x}\right)f+v\left(\frac{\partial f}{\partial t}+v\frac{\partial f}{\partial x}\right).\nonumber
\end{eqnarray}
Moreover, we define the action function $S(x,t)$ as
\begin{eqnarray}
\label{sec3-osc-re-s}
S(x,t)=F(x,t)+v(x,t)f(x,t),
\end{eqnarray}
then we have
\begin{eqnarray}
\label{sec3-osc-re-ft}
\frac{\partial F}{\partial t}=\frac{\partial S}{\partial t}-v\frac{\partial f}{\partial t}-\frac{\partial v}{\partial t}f,\\
\label{sec3-osc-re-fx}
\frac{\partial F}{\partial x}=\frac{\partial S}{\partial x}-v\frac{\partial f}{\partial x}-\frac{\partial v}{\partial x}f.
\end{eqnarray}
Using these two equations, Eq.~(\ref{sec3-cara-2nd-dx-sol-re}) becomes
\begin{eqnarray}
\label{sec3-cara-2nd-dx-sol-re-fx}
mv=\frac{\partial S}{\partial x}+\frac{\partial f}{\partial t}+v\frac{\partial f}{\partial x}-\frac{\partial v}{\partial x}f,
\end{eqnarray}
and (\ref{sec3-osc-re}) is simplified to
\begin{eqnarray}
\label{sec3-osc-fx}
\frac{\partial S}{\partial t}+v\frac{\partial S}{\partial x}=-\frac{m\epsilon}{2\omega^2}\left(\frac{\partial v}{\partial t}+v\frac{\partial v}{\partial x}\right)^2+\frac{1}{2}mv^2-V(x).
\end{eqnarray}
We can eliminate $f$ from Eq.~(\ref{sec3-cara-2nd-dx-sol-re-fx}) with its expression in Eq.~(\ref{sec3-cara-2nd-2dx-sol-re}), then we obtain
\begin{eqnarray}
\label{sec3-cara-2nd-dx-sol-f}
\frac{\partial S}{\partial x}=mv+\frac{m\epsilon}{\omega^2}\left(\frac{\partial^2 v}{\partial t^2}+2v\frac{\partial^2 v}{\partial t\partial x}+v^2\frac{\partial^2 v}{\partial x^2}\right).
\end{eqnarray}

Eqs.~(\ref{sec3-osc-fx}) and (\ref{sec3-cara-2nd-dx-sol-f}) are the novel Hamilton-Jacobi equations derived through Caratheodory's approach. They are the correspondences to Eqs.~(\ref{sec2-osc-ham-jac}) and (\ref{sec2-osc-ham-jac-p}), which are the canonical HJEs derived through Ostrogradsky's approach. In the derivation of Eqs.~(\ref{sec3-osc-fx}) and (\ref{sec3-cara-2nd-dx-sol-f}), we have not employed canonical variables. Instead, the velocity field $v(x,t)$ is the critical variable. We could name Eqs.~(\ref{sec3-osc-fx}) and (\ref{sec3-cara-2nd-dx-sol-f}) as the velocity field formalism of Hamilton-Jacobi equation. 

If we can solve the velocity field $v(x,t)$ in terms of $\frac{\partial S}{\partial x}$ through Eq.~(\ref{sec3-cara-2nd-dx-sol-f}), after inserting this solution of $v(x,t)$ into Eq.~(\ref{sec3-osc-fx}), then we will obtain a single equation correspondent to Eq.~(\ref{sec2-osc-ham-jac-form}). However, Eq.~(\ref{sec3-cara-2nd-dx-sol-f}) is a second-order and nonlinear partial differential equation of $v(x,t)$, which is difficult to find its general solution. In the below, we will take a look at its series solution.

Eq.~(\ref{sec3-osc-fx}) can be rewritten as
\begin{eqnarray}
\label{sec3-osc-fx-re}
\frac{\partial S}{\partial t}+\frac{1}{2m}\left(\frac{\partial S}{\partial x}\right)^2+V(x)=\frac{1}{2m}\left(\frac{\partial S}{\partial x}-mv\right)^2-\frac{m\epsilon}{2\omega^2}\left(\frac{\partial v}{\partial t}+v\frac{\partial v}{\partial x}\right)^2.
\end{eqnarray}
This formulation is useful for the series expansion about $\epsilon$. For $\epsilon=0$, we obtain from Eq.~(\ref{sec3-cara-2nd-dx-sol-f})
\begin{eqnarray}
\label{sec3-cara-2nd-dx-sol-f-v}
v_0=\frac{1}{m}\frac{\partial S}{\partial x}.
\end{eqnarray}
In this case, Eq.~(\ref{sec3-osc-fx-re}) reduces to the conventional HJE. For $\epsilon\neq{0}$,  we can derive a series solution about $\epsilon$ from Eq.~(\ref{sec3-cara-2nd-dx-sol-f})
\begin{eqnarray}
\label{sec3-cara-2nd-dx-sol-f-v-ser}
v=\frac{1}{m}\frac{\partial S}{\partial x}-\frac{\epsilon}{\omega^2}\left(\frac{\partial^3 S}{\partial t^2\partial x}+\frac{2}{m}\frac{\partial S}{\partial x}\frac{\partial^3 S}{\partial t\partial x^2}+\frac{1}{m^2}\left(\frac{\partial S}{\partial x}\right)^2\frac{\partial^3 S}{\partial x^3}\right)+\cdots.
\end{eqnarray}
Inserting this expression of $v$ into Eq.~(\ref{sec3-osc-fx-re}), we obtain a series expansion of Eq.~(\ref{sec3-osc-fx-re})
\begin{eqnarray}
\label{sec3-osc-fx-re-s}
\frac{\partial S}{\partial t}+H_0+\epsilon H_1+\epsilon^2 H_2+\cdots=0,
\end{eqnarray}
where $H_0$ is
\begin{eqnarray}
\label{sec3-osc-fx-re-sa}
H_0=\frac{1}{2m}\left(\frac{\partial S}{\partial x}\right)^2+V(x),
\end{eqnarray}
which is the conventional Hamiltonian.  $H_1$ is
\begin{eqnarray}
\label{sec3-osc-fx-re-sb}
H_1&=&\frac{1}{2m\omega^2}\left(\frac{\partial^2 S}{\partial t\partial x}+\frac{1}{m}\frac{\partial S}{\partial x}\frac{\partial^2 S}{\partial x}\right)^2\\
&=&\frac{1}{2m\omega^2}\left\{\left(\frac{\partial^2 S}{\partial t\partial x}\right)^2+\frac{2}{m}\frac{\partial^2 S}{\partial t\partial x}\frac{\partial S}{\partial x}\frac{\partial^2 S}{\partial x}+\frac{1}{m^2}\left(\frac{\partial S}{\partial x}\frac{\partial^2 S}{\partial x}\right)^2\right\}.\nonumber
\end{eqnarray}
Eq.~(\ref{sec3-osc-fx-re-s}) is a single differential equation about the action function $S(t,x)$, which is correspondent to Eq.~(\ref{sec2-osc-ham-jac-form}).

For the future discussions, we are going to provide another deformation of the Hamilton-Jacobi equation~(\ref{sec3-osc-fx}). We define the quantity
\begin{eqnarray}
\label{sec3-osc-fx-b-re}
u=v-\frac{1}{m}\frac{\partial S}{\partial x},
\end{eqnarray}
which yields
\begin{eqnarray}
\label{sec3-osc-fx-b-re-u}
v=u+\frac{1}{m}\frac{\partial S}{\partial x}.
\end{eqnarray}
We also define
\begin{eqnarray}
\label{sec3-osc-fx-b-re-w}
K=-\frac{\epsilon}{\omega^2}\left(\frac{\partial^2 v}{\partial t^2}+2v\frac{\partial^2 v}{\partial t\partial x}+v^2\frac{\partial^2 v}{\partial x^2}\right).
\end{eqnarray}
Then Eq.~(\ref{sec3-cara-2nd-dx-sol-f}) can be reformulated as
\begin{eqnarray}
\label{sec3-osc-fx-b-v}
u-K=0.
\end{eqnarray}
With the relation between $u$ and $v$ in Eq.~(\ref{sec3-osc-fx-b-re-u}), we obtain the identity
\begin{eqnarray}
\label{sec3-osc-fx-b-u}
\frac{\partial v}{\partial t}+v\frac{\partial v}{\partial x}&=&\frac{\partial}{\partial t}\left(u+\tfrac{1}{m}\tfrac{\partial S}{\partial x}\right)+\left(u+\tfrac{1}{m}\tfrac{\partial S}{\partial x}\right)\frac{\partial}{\partial x}\left(u+\tfrac{1}{m}\tfrac{\partial S}{\partial x}\right)\nonumber\\
&=&\frac{\partial u}{\partial t}+u\frac{\partial u}{\partial x}+\frac{1}{m}\frac{\partial}{\partial x}\left(M+u\tfrac{\partial S}{\partial x}\right),
\end{eqnarray}
where $M$ is given as
\begin{eqnarray}
\label{sec3-osc-fx-b-u-m}
M=\frac{\partial S}{\partial t}+\frac{1}{2m}\left(\frac{\partial S}{\partial x}\right)^2.
\end{eqnarray}
Considering Eq.~(\ref{sec3-osc-fx-b-v}), we can replace $u$ with $K$. Then Eq.~(\ref{sec3-osc-fx-b-u}) turns into
\begin{eqnarray}
\label{sec3-osc-fx-b-m}
\frac{\partial v}{\partial t}+v\frac{\partial v}{\partial x}
&=&\frac{\partial K}{\partial t}+K\frac{\partial K}{\partial x}+\frac{1}{m}\frac{\partial}{\partial x}\left(M+K\tfrac{\partial S}{\partial x}\right).
\end{eqnarray}
With the above discussions, Eq.~(\ref{sec3-osc-fx-re}) can be recast into the following new formulation
\begin{eqnarray}
\label{sec3-osc-fx-re-w-s}
\frac{\partial S}{\partial t}+\frac{1}{2m}\left(\frac{\partial S}{\partial x}\right)^2+V(x)=\frac{m}{2}K^2-\frac{m\epsilon}{2\omega^2}\left[\frac{\partial K}{\partial t}+K\frac{\partial K}{\partial x}+\frac{1}{m}\frac{\partial}{\partial x}\left(M+K\tfrac{\partial S}{\partial x}\right)\right]^2,
\end{eqnarray}
which is also
\begin{eqnarray}
\label{sec3-osc-fx-re-w}
M+V(x)=\frac{m}{2}K^2-\frac{m\epsilon}{2\omega^2}\left[\frac{\partial K}{\partial t}+K\frac{\partial K}{\partial x}+\frac{1}{m}\frac{\partial}{\partial x}\left(M+K\tfrac{\partial S}{\partial x}\right)\right]^2.
\end{eqnarray}
The expressions of $M$ and $K$ are defined in Eqs.~(\ref{sec3-osc-fx-b-u-m}) and (\ref{sec3-osc-fx-b-re-w}) respectively. In this deformed equation~(\ref{sec3-osc-fx-re-w}), the elemental variables are still the action $S$ and the velocity field $v$. 

\subsection{Classical Solution of Hamilton-Jacobi Equation}\label{sec3a}

As is known from the classical mechanics, the solution of Hamilton-Jacobi equation~(HJE) can be used to generate the solution of equation of motion~(EOM). In this subsection, our purpose is to verify that this feature is established for the novel HJEs~(\ref{sec3-osc-fx}) and (\ref{sec3-cara-2nd-dx-sol-f}). At first, we consider the solution of the following ansatz
\begin{eqnarray}
\label{sec3a-osc-hj-form-t}
S(x,t)=-Et+W(x),~~ v(x,t)=\xi(x).
\end{eqnarray}
Eq.~(\ref{sec3-osc-fx}) and (\ref{sec3-cara-2nd-dx-sol-f}) are computed as
\begin{eqnarray}
\label{sec3a-osc-fx}
-E+\xi\frac{dW}{dx}&=&-\frac{m\epsilon}{2\omega^2}\left(\xi\frac{d\xi}{dx}\right)^2+\frac{1}{2}m\xi^2-V(x),\\
\label{sec3b-cara-2nd-dx-sol-f}
\frac{dW}{dx}&=&m\xi+\frac{m\epsilon}{\omega^2}\xi^2\frac{d^2 \xi}{dx^2}.
\end{eqnarray}
$\frac{dW}{dx}$ can be eliminated from Eq.~(\ref{sec3a-osc-fx}), then we obtain
\begin{eqnarray}
\label{sec3a-osc-fx-w}
E-V(x)=\frac{1}{2}m\xi^2+\frac{m\epsilon}{2\omega^2}\left\{\left(\xi\frac{d\xi}{dx}\right)^2+2\xi^3\frac{d^2 \xi}{dx^2}\right\}.
\end{eqnarray}

In order to establish the relation between HJE and EOM, we employ Eqs.~(\ref{sec3-osc-cara-2nd-dx-sol-a}) and (\ref{sec3a-osc-hj-form-t}) to obtain
\begin{eqnarray}
\label{sec3a-osc-cara-2nd-dx-sol-a}
\dot{x}=v(x,t)=\xi(x).
\end{eqnarray}
Once we obtain the explicit solution of $\xi(x)$, then $x(t)$ can be solved from Eq.~(\ref{sec3a-osc-cara-2nd-dx-sol-a}). In general, we are going to verify that if $\xi(x)$ satisfies the HJE~(\ref{sec3a-osc-fx-w}), then $x(t)$ determined by Eq.~(\ref{sec3a-osc-cara-2nd-dx-sol-a}) will satisfy the EOM~(\ref{sec2-osc-el}). The verification is as follows. From Eq.~(\ref{sec3a-osc-cara-2nd-dx-sol-a}), we have
\begin{eqnarray}
\label{sec3a-osc-cara-2nd-dx-sol-a-2}
\ddot{x}&=&\frac{d\xi}{dt}=\frac{dx}{dt}\frac{d\xi}{dx}=\dot{x}\frac{d\xi}{dx}=\xi\frac{d\xi}{dx}.
\end{eqnarray}
Similar computations yield the equations
\begin{eqnarray}
\label{sec3a-osc-cara-2nd-dx-sol-a-3}
\dddot{x}&=&\xi\left(\frac{d\xi}{dx}\right)^2+\xi^2\frac{d^2\xi}{dx^2},\\
\label{sec3a-osc-cara-2nd-dx-sol-a-4}
{x}^{(4)}&=&\xi\left(\frac{d\xi}{dx}\right)^3+4\xi^2\frac{d\xi}{dx}\frac{d^2\xi}{dx^2}
+\xi^3\frac{d^3\xi}{dx^3}.
\end{eqnarray}
In order to check whether the EOM~(\ref{sec2-osc-el}) is satisfied, we need to check whether the quantity
\begin{eqnarray}
\label{sec3a-ver}
\frac{m\epsilon}{\omega^2}{x}^{(4)}+m\ddot{x}+\frac{dV}{dx}=\frac{m\epsilon}{\omega^2}\left[\xi\left(\frac{d\xi}{dx}\right)^3+4\xi^2\frac{d\xi}{dx}\frac{d^2\xi}{dx^2}
+\xi^3\frac{d^3\xi}{dx^3}\right]+m\xi\frac{d\xi}{dx}+\frac{dV}{dx}
\end{eqnarray}
is zero. From Eq.~(\ref{sec3a-osc-fx-w}), we have
\begin{eqnarray}
\label{sec3a-osc-fx-w-s}
\frac{d^2 \xi}{dx^2}=\frac{\omega^2}{m\epsilon}\frac{1}{\xi^3}\left[E-V(x)-\frac{1}{2}m\xi^2-\frac{m\epsilon}{2\omega^2}\left(\xi\frac{d\xi}{dx}\right)^2\right],
\end{eqnarray}
which differentiation yields
\begin{eqnarray}
\label{sec3a-osc-fx-w-s-3}
\frac{d^3 \xi}{dx^3}&=&-\frac{\omega^2}{m\epsilon}\frac{3}{\xi^4}\frac{d\xi}{dx}
\left[E-V(x)-\frac{1}{2}m\xi^2-\frac{m\epsilon}{2\omega^2}\left(\xi\frac{d\xi}{dx}\right)^2\right]\\
&+&\frac{\omega^2}{m\epsilon}\frac{1}{\xi^3}\left[-\frac{m\epsilon}{\omega^2}\left(\xi\frac{d\xi}{dx}\right)\left(\xi\frac{d^2\xi}{dx^2}+\frac{d\xi}{dx}\frac{d\xi}{dx}\right)-m\xi\frac{d\xi}{dx}-\frac{dV}{dx}\right].\nonumber
\end{eqnarray}
With Eq.~(\ref{sec3a-osc-fx-w-s}), $\frac{d^2\xi}{dx^2}$ can be eliminated from Eq.~(\ref{sec3a-osc-fx-w-s-3}), then we obtain
\begin{eqnarray}
\label{sec3a-osc-fx-w-s-3b}
\frac{d^3 \xi}{dx^3}&=&-\frac{\omega^2}{m\epsilon}\frac{4}{\xi^4}\frac{d\xi}{dx}
\left[E-\frac{\epsilon}{2}\frac{m}{\omega^2}\left(\xi\frac{d\xi}{dx}\right)^2-\frac{1}{2}m\xi^2-V(x)\right]\\
&+&\frac{\omega^2}{m\epsilon}\frac{1}{\xi^3}\left[-\frac{m\epsilon}{\omega^2}\xi\left(\frac{d\xi}{dx}\right)^3-m\xi\frac{d\xi}{dx}-\frac{dV}{dx}\right].\nonumber
\end{eqnarray}
Inserting the foregoing expressions of $\frac{d^2 \xi}{dx^2}$ and $\frac{d^3 \xi}{dx^3}$ into Eq.~(\ref{sec3a-ver}), we obtain the identity
\begin{eqnarray}
\label{sec3a-ver-b}
\frac{m\epsilon}{\omega^2}\left[\xi\left(\frac{d\xi}{dx}\right)^3+4\xi^2\frac{d\xi}{dx}\frac{d^2\xi}{dx^2}
+\xi^3\frac{d^3\xi}{dx^3}\right]+m\xi\frac{d\xi}{dx}+\frac{dV}{dx}=0,
\end{eqnarray}
which means that the EOM~(\ref{sec2-osc-el}) is satisfied. 

The above arguments have shown that the solutions of HJE can generate the solutions of EOM. We shall present two specific examples in the following two subsections.

\subsubsection{Free Particle Potential}\label{sec3aa}

For the free particle potential~(\ref{sec2aa-pot}), Eq.~(\ref{sec3a-osc-fx-w}) becomes
\begin{eqnarray}
\label{sec3aa-osc-fx}
E-\frac{1}{2}m\xi^2=\frac{m\epsilon}{2\omega^2}\left(\xi\frac{d\xi}{dx}\right)^2+\frac{m\epsilon}{\omega^2}\xi^3\frac{d^2 \xi}{dx^2}.
\end{eqnarray}
Multiplied by $\xi^{-2}\tfrac{d\xi}{dx}$ on both sides, Eq.~(\ref{sec3aa-osc-fx}) can be integrated once to yield
\begin{eqnarray}
\label{sec3aa-osc-fx-int}
\frac{\epsilon}{\omega^2}\frac{\xi}{2}\left(\frac{d\xi}{dx}\right)^2+\left(\frac{\xi}{2}+\frac{E}{m}\frac{1}{\xi}\right)=c_1,
\end{eqnarray}
where $c_1$ is an integral constant. This equation can be solved as the implicit function formalism
\begin{eqnarray}
\label{sec3aa-osc-fx-int-sol}
c_1 \arcsin\tfrac{1}{A}\left(\xi-c_1\right)-\sqrt{A^2-(\xi-c_1)^2}=\frac{\omega}{\sqrt{\epsilon}}x+c_2,
\end{eqnarray}
where $c_2$ is another integral constant, and $A$ has been defined by Eq.~(\ref{sec2aa-sol-a}). Because of the implicit function formalism, it is not straightforward to work out the solution of EOM. Instead, we consider two simpler cases. Two special solutions of Eq.~(\ref{sec3aa-osc-fx}) are given as
\begin{eqnarray}
\label{sec3aa-osc-fx-int-sa}
\xi_{+}(x)&=&\sqrt{\tfrac{2E}{m}},\\
\label{sec3aa-osc-fx-int-sb}
\xi_{-}(x)&=&\sqrt{-\tfrac{2E}{m}-\tfrac{\omega^2}{\epsilon}x^2}.
\end{eqnarray}
With Eqs.~(\ref{sec3a-osc-cara-2nd-dx-sol-a}), (\ref{sec3aa-osc-fx-int-sa}) and (\ref{sec3aa-osc-fx-int-sb}), we obtain
\begin{eqnarray}
\label{sec3a-osc-fx-w-sol-a-xi}
\frac{dx}{dt}&=&\xi_{+}(x)\Longrightarrow x_{+}(t)=\sqrt{\tfrac{2E}{m}}\left(t-t_{+}\right),\\
\label{sec3a-osc-fx-w-sol-b-xi}
\frac{dx}{dt}&=&\xi_{-}(x)\Longrightarrow x_{-}(t)=\tfrac{\sqrt{\epsilon}}{\omega}\sqrt{-\tfrac{2E}{m}}\sin\tfrac{\omega}{\sqrt{\epsilon}}(t-t_{-}),
\end{eqnarray}
where $t_{+}$ and $t_{-}$ are integral constants. These are two special solutions of EOM. Therefore, the special solutions of HJE generate two special cases of the general solution~(\ref{sec2aa-sol-x}).

\subsubsection{Harmonic Oscillator Potential}\label{sec3ab}

For the harmonic oscillator potential~(\ref{sec2ab-pot}), Eq.~(\ref{sec3a-osc-fx-w}) turns into
\begin{eqnarray}
\label{sec3ab-osc-fx}
E-\frac{1}{2}m\omega^2 x^2=\frac{1}{2}m\xi^2+\frac{m\epsilon}{2\omega^2}
\left\{\left(\xi\frac{d\xi}{dx}\right)^2+2\xi^3\frac{d^2 \xi}{dx^2}\right\}.
\end{eqnarray}
This is a nonlinear ordinary differential equation, which is difficult to find its general solution. Two special solutions are given as
\begin{eqnarray}
\label{sec3a-osc-fx-w-sol-a}
\xi_{+}(x)&=&\left[-\tfrac{2E}{m\sqrt{1-4\epsilon}}-\omega_{+}^2 x^2\right]^{\frac{1}{2}}
\\
\label{sec3a-osc-fx-w-sol-b}
\xi_{-}(x)&=&\left[\tfrac{2E}{m\sqrt{1-4\epsilon}}-\omega_{-}^2 x^2\right]^{\frac{1}{2}}.
\end{eqnarray}
With Eqs.~(\ref{sec3a-osc-cara-2nd-dx-sol-a}), (\ref{sec3a-osc-fx-w-sol-a}) and (\ref{sec3a-osc-fx-w-sol-b}), we obtain
\begin{eqnarray}
\label{sec3a-osc-fx-w-sol-a-xi}
\frac{dx}{dt}&=&\xi_{+}(x)\Longrightarrow x_{+}(t)=\tfrac{1}{\omega_{+}}\left[-\tfrac{2E}{m\sqrt{1-4\epsilon}}\right]^{\frac{1}{2}}\sin{\omega_{+}(t-t_{+})},\\
\label{sec3a-osc-fx-w-sol-b-xi}
\frac{dx}{dt}&=&\xi_{-}(x)\Longrightarrow x_{-}(t)=\tfrac{1}{\omega_{-}}\left[\tfrac{2E}{m\sqrt{1-4\epsilon}}\right]^{\frac{1}{2}}\sin{\omega_{-}(t-t_{-})},
\end{eqnarray}
where $t_{+}$ and $t_{-}$ are integral constants. These solutions are special cases of the general solution in Eq.~(\ref{sec2-osc-el-sol}). Therefore, $\xi_{+}(x)$ and $\xi_{-}(x)$ generate two special solutions of the EOM~(\ref{sec2-osc-el}) respectively.

The ansatz~(\ref{sec3a-osc-hj-form-t}) can not be solved directly in the present situation. As an alternative, Eqs.~(\ref{sec3-osc-fx}) and (\ref{sec3-cara-2nd-dx-sol-f}) can be solved by the polynomial ansatz
\begin{eqnarray}
\label{sec3a-osc-hj-form-s}
S(t,x)&=&\alpha(t)x^2+\beta(t)x+\chi(t),\\
\label{sec3a-osc-hj-form-v}
v(t,x)&=&\kappa(t)x+\sigma(t).
\end{eqnarray}
The coefficients of $x$ related terms in Eq.~(\ref{sec3-osc-fx}) yield the equations
\begin{eqnarray}
\label{sec3a-osc-hj-form-s-xa}
\dot{\chi}+\beta\sigma-\frac{m}{2}\sigma^2
+\frac{m\epsilon}{2\omega^2}\left(\dot{\sigma}+\kappa\sigma\right)^2&=&0,\\
\label{sec3a-osc-hj-form-s-xb}
\dot{\beta}+2\alpha\sigma+\beta\kappa-m\kappa\sigma
+\frac{m\epsilon}{\omega^2}\left(\dot{\sigma}+\kappa\sigma\right)\left(\dot{\kappa}+\kappa^2\right)&=&0,\\
\label{sec3a-osc-hj-form-s-xc}
\dot{\alpha}+2\alpha\kappa-\frac{m}{2}\kappa^2+\frac{1}{2}m\omega^2
+\frac{m\epsilon}{2\omega^2}\left(\dot{\kappa}+\kappa^2\right)^2&=&0,
\end{eqnarray}
and the coefficients of Eq.~(\ref{sec3-cara-2nd-dx-sol-f}) yield the equations
\begin{eqnarray}
\label{sec3a-osc-hj-form-v-xa}
\frac{m\epsilon}{\omega^2}\left(\ddot{\sigma}+2\sigma\dot{\kappa}\right)
-\beta+m\sigma&=&0,\\
\label{sec3a-osc-hj-form-v-xb}
\frac{m\epsilon}{\omega^2}\left(\ddot{\kappa}+2\kappa\dot{\kappa}\right)
-2\alpha+m\kappa&=&0.
\end{eqnarray}
From Eq.~(\ref{sec3a-osc-hj-form-v-xb}), we solve $\alpha(t)$ as
\begin{eqnarray}
\label{sec3a-osc-hj-form-v-xb-sol}
\alpha=\frac{m\epsilon}{2\omega^2}\left(\ddot{\kappa}+2\kappa\dot{\kappa}\right)
+\frac{1}{2}m\kappa.
\end{eqnarray}
Inserting this expression of $\alpha(t)$ into Eq.~(\ref{sec3a-osc-hj-form-s-xc}), we obtain the single equation of $\kappa(t)$
\begin{eqnarray}
\label{sec3a-osc-hj-form-s-v}
\frac{\epsilon}{\omega^2}\left(\dddot{\kappa}+3\dot{\kappa}^2+4\kappa\ddot{\kappa}+6\kappa^2\dot{\kappa}+\kappa^4\right)+\dot{\kappa}+\kappa^2+\omega^2=0.
\end{eqnarray}
Using the transformation
\begin{eqnarray}
\label{sec3a-osc-hj-form-svt}
\kappa(t)=\frac{\dot{\eta}}{\eta},
\end{eqnarray}
Eq.~(\ref{sec3a-osc-hj-form-s-v}) becomes
\begin{eqnarray}
\label{sec3a-osc-hj-form-sv-t}
\frac{\epsilon}{\omega^2}\eta^{(4)}+\ddot{\eta}+\omega^2\eta=0.
\end{eqnarray}
This is a linear ordinary differential equation, which coincides with the EOM~(\ref{sec2-osc-el}) for the harmonic oscillator potential~(\ref{sec2ab-pot}). Its solution of $\eta(t)$ can be given as
\begin{eqnarray}
\label{sec3a-osc-hj-form-sv-t-sol}
\eta(t)=\eta_{+}\cos[\omega_{+}(t-t_+)]+\eta_{-}\cos[\omega_{-}(t-t_{-})],
\end{eqnarray}
where $\eta_{\pm}$ are integral constants. This solution is equivalent to the formulation in Eq.~(\ref{sec2-osc-el-sol}). Using $\eta(t)$, we can obtain the solution of $\kappa(t)$ from Eq.~(\ref{sec3a-osc-hj-form-svt}). 
From Eq.~(\ref{sec3a-osc-hj-form-s-xb}) and (\ref{sec3a-osc-hj-form-v-xa}), we obtain
the equation of $\sigma(t)$
\begin{eqnarray}
\label{sec3a-osc-hj-form-sv-s}
\frac{\epsilon}{\omega^2}\left(\dddot{\sigma}+3\dot{\kappa}\dot{\sigma}+3\sigma\ddot{\kappa}+\kappa\ddot{\sigma}+5\sigma\kappa\dot{\kappa}+\kappa^2\dot{\sigma}+\kappa^3\sigma\right)+\dot{\sigma}+\kappa\sigma=0.
\end{eqnarray}
Using the transformation
\begin{eqnarray}
\label{sec3a-osc-hj-form-svs}
\sigma(t)=\frac{d}{dt}\varsigma(t)-\kappa(t)\varsigma(t),
\end{eqnarray}
Eq.~(\ref{sec3a-osc-hj-form-sv-s}) turns into
\begin{eqnarray}
\label{sec3a-osc-hj-form-sv-vs}
\frac{\epsilon}{\omega^2}\left(\varsigma^{(4)}-\left(\dddot{\kappa}+3\dot{\kappa}^2+4\kappa\ddot{\kappa}+6\kappa^2\dot{\kappa}+\kappa^4\right)\varsigma\right)+\ddot{\varsigma}-(\dot{\kappa}+\kappa^2)\varsigma=0.
\end{eqnarray}
Using Eq.~(\ref{sec3a-osc-hj-form-s-v}) about $\kappa(t)$, Eq.~(\ref{sec3a-osc-hj-form-sv-vs}) becomes
\begin{eqnarray}
\label{sec3a-osc-hj-form-svvs}
\frac{\epsilon}{\omega^2}\varsigma^{(4)}+\ddot{\varsigma}+\omega^2\varsigma=0.
\end{eqnarray}
The solution $\varsigma(t)$ has the same formulation as Eq.~(\ref{sec3a-osc-hj-form-sv-t-sol})
\begin{eqnarray}
\label{sec3a-osc-hj-form-sv-t-sol-si}
\varsigma(t)=\varsigma_{+}\cos[\omega_{+}(t-z_+)]+\varsigma_{-}\cos[\omega_{-}(t-z_{-})],
\end{eqnarray}
where $\varsigma_{\pm}$ and $z_{\pm}$ are integral constants. Using Eqs.~(\ref{sec3a-osc-hj-form-svt}), (\ref{sec3a-osc-hj-form-svs}), (\ref{sec3a-osc-hj-form-v-xa}), (\ref{sec3a-osc-hj-form-v-xb}) and (\ref{sec3a-osc-hj-form-s-xa}), the solutions of $\kappa(t)$, $\sigma(t)$, $\alpha(t)$, $\beta(t)$ and $\chi(t)$ can be determined by $\eta(t)$ and $\varsigma(t)$ correspondingly. 

In the above, we have derived the solution of HJE with the polynomial ansatz~(\ref{sec3a-osc-hj-form-s}) and (\ref{sec3a-osc-hj-form-v}). In order to generate the solution of EOM~(\ref{sec2-osc-el}), we consider Eqs.~(\ref{sec3-osc-cara-2nd-dx-sol-a}) and (\ref{sec3a-osc-hj-form-v})
\begin{eqnarray}
\label{sec3b-osc-cara-2nd-dx-sol-a}
\dot{x}=v(x,t)=\kappa(t)x+\sigma(t).
\end{eqnarray}
With the transformation in Eq.~(\ref{sec3a-osc-hj-form-svs}), we have
\begin{eqnarray}
\label{sec3b-osc-cara-x}
\dot{x}-\dot{\varsigma}=\kappa(t)(x-\varsigma).
\end{eqnarray}
One special solution of this equation is
\begin{eqnarray}
\label{sec3b-osc-cara-x-sol}
x(t)=\varsigma(t),
\end{eqnarray}
which means that $x(t)$ satisfies Eq.~(\ref{sec3a-osc-hj-form-svvs}). Therefore, the polynomial solution in the formulation of Eqs.~(\ref{sec3a-osc-hj-form-s}) and (\ref{sec3a-osc-hj-form-v}) can be used to generate the solution of EOM~(\ref{sec2-osc-el}).

\subsection{Quantum mechanics}\label{sec3b}

Eqs.~(\ref{sec3-osc-fx-re-w}) and (\ref{sec3-cara-2nd-dx-sol-f}) are the derived novel formulations of Hamilton-Jacobi equations. These equations do not possess the canonical structure or symplectic structure~\cite{Faddeev:1988qp}. Therefore, it is extremely difficult to obtain quantum mechanical correspondences for these equations straightforwardly. One point of view could be that Eqs.~(\ref{sec3-osc-fx-re-w}) and (\ref{sec3-cara-2nd-dx-sol-f}) are not in the suitable formulations for quantum mechanical treatments. However, as shown in section~\ref{sec2b}, there is a close relation between the Schr$\ddot{\mathrm{o}}$dinger equation and the Hamilton-Jacobi equation. Thanks to this relation, we may attempt to seek some Schr$\ddot{\mathrm{o}}$dinger-type equation, which can produce the novel Hamilton-Jacobi equation~(\ref{sec3-osc-fx-re-w}) in the classical limit $\hbar\rightarrow 0$.

We rewrite the wave function $\Psi$ as
\begin{eqnarray}
\label{sec3b-osc-psi}
\Psi(x,t)=R(x,t) e^{\frac{i}{\hbar}S(x,t)},
\end{eqnarray}
then we have the computations
\begin{eqnarray}
\label{sec3b-osc-psi-x}
-i\hbar\frac{1}{\Psi} \frac{\partial \Psi}{\partial x}&=&\frac{\partial S}{\partial x}-i\hbar\frac{1}{R} \frac{\partial R}{\partial x},\\
\label{sec3b-osc-psi-xx}
-\hbar^2\frac{1}{\Psi} \frac{\partial^2 \Psi}{\partial x^2}&=&\left(\frac{\partial S}{\partial x}\right)^2-\hbar^2\frac{1}{R} \frac{\partial^2 R}{\partial x^2}-i\hbar\frac{\partial^2 S}{\partial x^2}-2i\hbar\frac{1}{R} \frac{\partial R}{\partial x}\frac{\partial S}{\partial x}.
\end{eqnarray}
From Eqs.~(\ref{sec3b-osc-psi-x}) and (\ref{sec3b-osc-psi-xx}), we obtain 
\begin{eqnarray}
\label{sec3b-psi-ss}
-\frac{i\hbar}{\Psi}\frac{\partial \Psi}{\partial x}\xrightarrow{\hbar=0}\frac{\partial S}{\partial x},~-\frac{\hbar^2}{\Psi}\frac{\partial^2 \Psi}{\partial x^2}\xrightarrow{\hbar=0}\left(\frac{\partial S}{\partial x}\right)^2.
\end{eqnarray}
These relations mean that the left-handed terms of these two formula are approximate to the right-handed terms in the limit $\hbar\rightarrow 0$. As another example, we consider the first term of the second line in Eq.~(\ref{sec3-osc-fx-re-sb}). Using Leibniz rules, we have the identity
\begin{eqnarray}
\label{sec3b-psi-ss-s}
\left(\frac{\partial^2 S}{\partial t\partial x}\right)^2=\frac{1}{2}\frac{\partial^2 }{\partial t^2}\left(\frac{\partial S}{\partial x}\right)^2-\frac{\partial S}{\partial x}\frac{\partial^2 }{\partial t^2}\frac{\partial S}{\partial x}.
\end{eqnarray}
With this identity and Eq.~(\ref{sec3b-psi-ss}), we have the following approximate formula
\begin{eqnarray}
\label{sec3b-psi-ss-sp}
\left\{-\frac{\hbar^2}{2}\frac{\partial^2 }{\partial t^2}\left(\frac{1}{\Psi}\frac{\partial^2 \Psi}{\partial x^2}\right)+\hbar^2\left(\frac{1}{\Psi}\frac{\partial \Psi}{\partial x}\right)\frac{\partial^2 }{\partial t^2}\left(\frac{1}{\Psi}\frac{\partial \Psi}{\partial x}\right)\right\}\xrightarrow{\hbar=0}\left(\frac{\partial^2 S}{\partial t\partial x}\right)^2,
\end{eqnarray}
which is a complicated variant of the formula in Eq.~(\ref{sec3b-psi-ss}). This equation produces the first term in Eq.~(\ref{sec3-osc-fx-re-sb}). For the other terms in Eq.~(\ref{sec3-osc-fx-re-sb}), we can also find the corresponding formula. There is a different feature between Eq.~(\ref{sec3b-psi-ss}) and Eq.~(\ref{sec3b-psi-ss-sp}). Multiplied by $\Psi$, the left-handed terms of Eq.~(\ref{sec3b-psi-ss}) are linear functions of $\Psi$; While the left-handed terms of Eq.~(\ref{sec3b-psi-ss-sp}) are nonlinear functions of $\Psi$. This observation implies that the underlying Schr$\ddot{\mathrm{o}}$dinger-type equations should be nonlinear equations.

In order to construct Schr$\ddot{\mathrm{o}}$dinger-type equations for Eq.~(\ref{sec3-osc-fx-re-w}), we should reverse the foregoing relations in Eq.~(\ref{sec3b-psi-ss}). We make the replacements
\begin{eqnarray}
\label{sec3b-osc-v}
\frac{\partial S}{\partial t}\rightarrow\hbar\mathbf{Im}\left(\frac{1}{\Psi}\frac{\partial\Psi}{\partial t}\right),~
\frac{\partial S}{\partial x}\rightarrow\hbar\mathbf{Im}\left(\frac{1}{\Psi}\frac{\partial\Psi}{\partial x}\right).
\end{eqnarray}
and
\begin{eqnarray}
\label{sec3b-osc-vv}
\left(\frac{\partial S}{\partial x}\right)^2\rightarrow-\hbar^2\mathbf{Re}\left(\frac{1}{\Psi}\frac{\partial^2\Psi}{\partial x^2}\right),~~\rho\rightarrow \Psi\Psi^{*}.
\end{eqnarray}
As already shown in Eqs.~(\ref{sec3b-osc-psi-x}), (\ref{sec3b-osc-psi-xx}) and (\ref{sec3b-psi-ss}), we can directly verify that the right-handed objects of Eqs.~(\ref{sec3b-osc-v}) and (\ref{sec3b-osc-vv}) generate the left-handed objects when $\hbar\rightarrow 0$.

Using the correspondent formula in Eqs.~(\ref{sec3b-osc-v}) and (\ref{sec3b-osc-vv}), we obtain the Schr$\ddot{\mathrm{o}}$dinger-type equation for 
Eq.~(\ref{sec3-osc-fx-re-w})
\begin{eqnarray}
\label{sec3b-osc-qm}
\mathbf{M}+V(x)
=\frac{1}{2}mK^2-\frac{m\epsilon}{2\omega^2}\left[\frac{\partial K}{\partial t}+K\frac{\partial K}{\partial x}+\frac{1}{m}\frac{\partial}{\partial x}\biggl(\mathbf{M}+\hbar K\mathbf{Im}\bigl(\tfrac{1}{\Psi}\tfrac{\partial\Psi}{\partial x}\bigr)\biggr)\right]^2,
\end{eqnarray}
where $\mathbf{M}$ is
\begin{eqnarray}
\label{sec3b-osc-qm-bm}
\mathbf{M}=\hbar\mathbf{Im}\left(\tfrac{1}{\Psi}\tfrac{\partial\Psi}{\partial t}\right)&-&\tfrac{\hbar^2}{2m}\mathbf{Re}\bigl(\tfrac{1}{\Psi}\tfrac{\partial^2\Psi}{\partial x^2}\bigr).
\end{eqnarray}
In these equations, $K$ is defined by Eq.~(\ref{sec3-osc-fx-b-re-w}). The corresponding equation for Eq.~(\ref{sec3-cara-2nd-dx-sol-f}) is
\begin{eqnarray}
\label{sec3b-cara-dx-sol-f}
\hbar\mathbf{Im}\left(\tfrac{1}{\Psi}\tfrac{\partial\Psi}{\partial x}\right)=mv+\frac{m\epsilon}{\omega^2}\left(\frac{\partial^2 v}{\partial t^2}+2v\frac{\partial^2 v}{\partial t\partial x}+v^2\frac{\partial^2 v}{\partial x^2}\right),
\end{eqnarray}
In the above, the terms associated with $S(x,t)$ are replaced by $\Psi$. However, the velocity field $v(t,x)$ remains unchanged. Through straightforward computations, we can verify that Eqs.~(\ref{sec3b-osc-qm}) and (\ref{sec3b-cara-dx-sol-f}) are approximate to Eqs.~(\ref{sec3-osc-fx-re-w}) and (\ref{sec3-cara-2nd-dx-sol-f}) in the limit $\hbar\rightarrow 0$. In other words, Eqs.~(\ref{sec3b-osc-qm-bm}) and (\ref{sec3b-cara-dx-sol-f}) are equivalent to Eqs.~(\ref{sec3-osc-fx-re-w}) and (\ref{sec3-cara-2nd-dx-sol-f}) plus quantum corrections which are proportional to $\hbar$. 

Eq.~(\ref{sec3b-osc-qm}) corresponds to the real part of the Schr$\ddot{\mathrm{o}}$dinger-type equation~(\ref{sec2-osc-sch-real}), which is not enough to determine $\Psi$ completely, because $\Psi$ is a complex function. For the consideration of quantum mechanics, we also have the conservation equation of probability density or the continuity equation
\begin{eqnarray}
\label{sec3b-con-p}
\frac{\partial \rho}{\partial t}+\frac{\partial}{\partial x}(\rho v)=0,
\end{eqnarray}
where $v(t,x)$ is the velocity field. Using the correspondence in Eq.~(\ref{sec3b-osc-vv}), we obtain
\begin{eqnarray}
\label{sec3b-con-p-r}
\frac{\partial}{\partial t}(\Psi\Psi^{*})+\frac{\partial}{\partial x}\left[(\Psi\Psi^{*})v\right]=0.
\end{eqnarray}
This equation is the corresponding formulation of Eq.~(\ref{sec2-osc-sch-ima-a-b}).

Eqs.~(\ref{sec3b-osc-qm}), (\ref{sec3b-cara-dx-sol-f}) and (\ref{sec3b-con-p-r}) are the derived Schr\"{o}dinger-type equations associated with the Hamilton-Jacobi equations~(\ref{sec3-osc-fx-re-w}) and (\ref{sec3-cara-2nd-dx-sol-f}). Eqs.~(\ref{sec3b-osc-qm}) and (\ref{sec3b-con-p-r}) are used to determine the complex wave functions $\Psi$. Eq.~(\ref{sec3b-cara-dx-sol-f}) can be interpreted as the equation to determine the velocity field $v$. For $\epsilon=0$, we obtain from Eq.~(\ref{sec3b-cara-dx-sol-f})
\begin{eqnarray}
\label{sec3b-con-p}
v_0=\tfrac{\hbar}{m}\mathbf{Im}\left(\tfrac{1}{\Psi}\tfrac{\partial\Psi}{\partial x}\right),
\end{eqnarray}
which is just the conventional velocity of probability current. In this situation, the higher derivative effects vanish, then Eq.~(\ref{sec3b-osc-qm}) and (\ref{sec3b-con-p-r}) are respectively reduced to be the real part and imaginary part of the conventional Schr\"{o}dinger equation.

Having derived the Schr\"{o}dinger-type equation, we are ready to seek the stationary solutions which have the following formulations 
\begin{eqnarray}
\label{sec3b-cara-sta}
\Psi(t,x)=\psi(x)e^{\frac{i}{\hbar}\theta(x)}e^{-\frac{i}{\hbar}Et},~~v(t,x)=\chi(x),
\end{eqnarray}
where we suppose that $\chi(x)$, $\psi(x)$ and $\theta(x)$ are real functions. With this ansatz, Eq.~(\ref{sec3b-con-p-r}) is computed as
\begin{eqnarray}
\label{sec3b-con-p-r-the}
\frac{d}{dx}\left[\psi(x)^{2}\chi(x)\right]=0,
\end{eqnarray}
which is solved as
\begin{eqnarray}
\label{sec3b-con-p-r-the-sol}
\chi(x)=\frac{\delta}{\psi(x)^{2}}.
\end{eqnarray}
Here $\delta$ is a integral constant. $K(t,x)$ and $\mathbf{M}$ are given as
\begin{eqnarray}
\label{sec3b-con-w}
K&=&-\frac{\epsilon}{\omega^2}\chi^2\frac{d^2\chi}{dx^2},\\
\label{sec3b-con-w-m}
\mathbf{M}&=&-E-\frac{\hbar^2}{2m}\frac{1}{\psi}\frac{d^2\psi}{dx^2}+\frac{1}{2m}\left(\frac{d\theta}{dx}\right)^2.
\end{eqnarray}
Eq.~(\ref{sec3b-osc-qm}) is computed as
\begin{eqnarray}
\label{sec3b-osc-qm-sta}
\mathbf{M}+V(x)=\frac{m}{2}K^2-\frac{m\epsilon}{2\omega^2}\left[K\frac{\partial K}{\partial x}+\frac{1}{m}\frac{\partial}{\partial x}\biggl(\mathbf{M}+K\tfrac{\partial\theta}{\partial x}\biggr)\right]^2.
\end{eqnarray}
Eq.~(\ref{sec3b-cara-dx-sol-f}) turns into
\begin{eqnarray}
\label{sec3b-dx-sol-f-r}
\frac{d\theta}{dx}=m\chi+\frac{m\epsilon}{\omega^2}\chi^2\frac{d^2\chi}{dx^2}.
\end{eqnarray}
For the integral constant $\delta\neq 0$, it is difficult to solve Eq.~(\ref{sec3b-osc-qm-sta}). For $\delta=0$, we have
\begin{eqnarray}
\label{sec3b-con-p-r-the-d}
\chi(x)=0.
\end{eqnarray}
In this case, Eq.~(\ref{sec3b-osc-qm-sta}) is equivalent to the following two equations
\begin{eqnarray}
\label{sec3b-osc-qm-sta-a}
-V(x)+U(x)
-\frac{\epsilon}{2m\omega^2}\left(\frac{dU}{d x}\right)^2=0,\\
\label{sec3b-osc-qm-sta-v}
-\frac{\hbar^2}{2m}\frac{1}{\psi}\frac{d^2\psi}{d x^2}=E-U(x).
\end{eqnarray}
We observed that Eq.~(\ref{sec3b-osc-qm-sta-v}) is the stationary Schr$\ddot{\mathrm{o}}$dinger equation in one dimensional space, which potential $U(x)$ is determined by Eq.~(\ref{sec3b-osc-qm-sta-a}). Therefore, in the specific case $\delta=0$, Eq.~(\ref{sec3b-osc-qm-sta}) can be interpreted as the conventional Schr$\ddot{\mathrm{o}}$dinger equation together with its potential determined by an ordinary differential equation.

From the above, we have seen that the Schr$\ddot{\mathrm{o}}$dinger-type equation~(\ref{sec3b-osc-qm}) is nonlinear, which is caused by the higher derivative terms in the Lagrangian. Generalized nonlinear Schr$\ddot{\mathrm{o}}$dinger equations have been considered from kinds of different backgrounds. Cubic and logarithmic wave mechanics were introduced in Refs.~\cite{Mielnik:1974hw,BialynickiBirula:1976zp}. The nonlinear Newton-Schr$\ddot{\mathrm{o}}$dinger equation was proposed to resolve the collapse problem of wave function~\cite{Penrose:1996cv,Bassi:2012bg}. Possible nonlinear corrections to quantum mechanics were constructed in Ref.~\cite{Weinberg:1989cm} and stringent limits were imposed by experimental tests~\cite{Bollinger:1989zz,Walsworth:1990zz,Majumder:1990zz}.  Besides, nonlinear quantum field theories were analyzed in~\cite{Kibble:1978vm}.

\subsubsection{Free Particle Potential}\label{sec3ba}

For the free particle potential in Eq.~(\ref{sec2aa-pot}), Eq.~(\ref{sec3b-osc-qm-sta-a}) turns into
\begin{eqnarray}
\label{sec3ba-osc-qm-sta-a}
U(x)-\frac{\epsilon}{2m\omega^2}\left(\frac{dU}{d x}\right)^2=0.
\end{eqnarray}
In this case, we can derive the exact solutions of Eq.~(\ref{sec3ba-osc-qm-sta-a}). There are two exact solutions. The first solution is
\begin{eqnarray}
\label{sec3ba-u-sol-va}
U_{o}(x)=0.
\end{eqnarray}
For this potential, Eq.~(\ref{sec3b-osc-qm-sta-v}) has the solution
\begin{eqnarray}
\label{sec3ba-qm-sol-va}
\psi_{o}(x)=c_1 \sin(kx)+c_2 \cos(kx), ~~E_{o}=\frac{\hbar^2 k^2}{2m}.
\end{eqnarray}
In the above, $c_1$ and $c_2$ are integral constants. This wave function $\psi_{o}(x)$ is understood as the wave function of free particle, and $E_{o}$ is the corresponding eigenenergy. The second exact solution of Eq.~(\ref{sec3ba-osc-qm-sta-a}) is
\begin{eqnarray}
\label{sec3ba-u-sol-vb}
U_{\epsilon}(x)=\frac{m}{2}\frac{\omega^2}{\epsilon}(x-x_0)^2,
\end{eqnarray}
where $x_0$ is a integral constant. For the potential $U_{\epsilon}$, Eq.~(\ref{sec3b-osc-qm-sta-v}) is equivalent to the Schr\"{o}dinger equation of the harmonic oscillator with frequency 
$\frac{\omega}{\sqrt{\epsilon}}$, which has the eigensolutions and eigenenergies
\begin{eqnarray}
\label{sec3ba-qm-sol-vb}
\psi_{n}(x)=A_n\exp\left(-\tfrac{m\omega}{2\hbar\sqrt{\epsilon}}(x-x_0)^2\right)H_{n}\left(\sqrt{\tfrac{m\omega}{\hbar\sqrt{\epsilon}}}(x-x_0)\right), ~~E_{n}=\left(n+\frac{1}{2}\right)\frac{\hbar\omega}{\sqrt{\epsilon}},
\end{eqnarray}
where $H_n(x)$ are the Hermite polynomials and $A_n$ are the normalized constants. 

The eigenenergies in Eqs.~(\ref{sec3ba-qm-sol-va}) and (\ref{sec3ba-qm-sol-vb}) are both positive. These results can be compared with that in subsection~\ref{sec2ba}, where the harmonic oscillator mode produces negative contributions to the eigenenergy. 

The potentials $U_o$ and $U_{\epsilon}$ have exhausted all the solutions of  Eq.~(\ref{sec3ba-osc-qm-sta-a}). Therefore, in the case of the free particle potential, there are not unbounded negative energy solutions. 

The forgoing discussions are based on the assumption $\delta=0$. For the case $\delta\neq 0$, because $\chi(x)$ is inversely proportional to the wave function $\psi(x)$ according to Eq.~(\ref{sec3b-con-p-r-the-sol}), the $\delta$ corrected wave function will be divergent at infinity if we use the wave functions $\psi_{n}(x)$ in Eq.~(\ref{sec3ba-qm-sol-vb}) as the first order approximations. This qualitative and perturbative observation implies that the non-zero $\delta$ corrections lead to physically unacceptable wave functions.

\subsubsection{Harmonic Oscillator Potential}\label{sec3bb}

For the harmonic oscillator potential in Eq.~(\ref{sec2ab-pot}), we obtain
\begin{eqnarray}
\label{sec3bb-osc-qm-sta-a}
-\frac{1}{2}m\omega x^2+U(x)-\frac{\epsilon}{2m\omega^2}\left(\frac{dU}{d x}\right)^2=0.
\end{eqnarray}
In this case, Eq.~(\ref{sec3bb-osc-qm-sta-a}) is a nonlinear ordinary differential equation, which is a special case of  Chrystal's equation. Its solutions can be classified into two types. 

The first type of solution is the polynomial solution. One polynomial solution of Eq.~(\ref{sec3bb-osc-qm-sta-a}) is given as
\begin{eqnarray}
\label{sec3bb-u-sol-va}
U_{+}(x)=\frac{1}{2}m\omega^2_{+}x^2,
\end{eqnarray}
where $\omega_{\pm}$ are defined by Eq.~(\ref{sec2-osc-el-sol-ome}). This is the harmonic oscillator potential with the frequency $\omega_{+}$. The eigenfunctions and eigenenergies of  Eq.~(\ref{sec3b-osc-qm-sta-v}) associated with this potential are
\begin{eqnarray}
\label{sec3bb-qm-sol-p}
\psi^{+}_{n}(x)=A^{+}_n\exp\left(-\tfrac{m\omega_{+}}{2\hbar}x^2\right)H_{n}\left(\sqrt{\tfrac{m\omega_{+}}{\hbar}}x\right), ~~E^{+}_{n}=\left(n+\frac{1}{2}\right)\hbar\omega_{+}.
\end{eqnarray}
Besides $U_{+}(x)$, Eq.~(\ref{sec3bb-osc-qm-sta-a}) has another polynomial solution 
$U_{-}(x)$
\begin{eqnarray}
\label{sec3bb-u-sol-vb}
U_{-}(x)=\frac{1}{2}m\omega^2_{-}x^2.
\end{eqnarray}
The eigenfuntions and eigenenergies for $U_{-}(x)$ are
\begin{eqnarray}
\label{sec3bb-qm-sol-m}
\psi^{-}_{n}(x)=A^{-}_n\exp\left(-\tfrac{m\omega_{-}}{2\hbar}x^2\right)H_{n}\left(\sqrt{\tfrac{m\omega_{-}}{\hbar}}x\right), ~~E^{-}_{n}=\left(n+\frac{1}{2}\right)\hbar\omega_{-}.
\end{eqnarray}
In the above, $A^{\pm}_n$ are the normalized constants. 

All the above solutions have the positive eigenenergies. In contrast with the result in Eq.~(\ref{sec2bb-sch-sol-a-c}), the $\omega_{+}$ related modes generate unbounded negative eigenenergies. 

The second type of solution is the general and exact solution, which can be derived as follows. Using the transformation
\begin{eqnarray}
\label{sec3bb-u-tr}
U(x)=\frac{1}{2}m\omega^2\left(1+\epsilon\cdot\zeta(x)^2\right)x^2,
\end{eqnarray}
We obtain from Eq.~(\ref{sec3bb-osc-qm-sta-a})
\begin{eqnarray}
\label{sec3bb-u-sol-tr}
\frac{\zeta d\zeta}{(\zeta-\mu_{+})(\zeta-\mu_{-})}=-\frac{dx}{x},
\end{eqnarray}
where
\begin{eqnarray}
\label{sec3bb-u-sol-tr-mu}
\mu_{\pm}=\frac{1\pm\sqrt{1-4\epsilon}}{2\epsilon}.
\end{eqnarray}
For $0<\epsilon\leq\frac{1}{4}$, we have $\mu_{+}\geq\mu_{-}$. For $\epsilon=\frac{1}{4}$, Eq.~(\ref{sec3bb-u-sol-tr}) is solved as
\begin{eqnarray}
\label{sec3bb-u-sol-tr-a}
\log\lvert\zeta-2\rvert-\frac{2}{\zeta-2}=-\log\lvert\tfrac{x}{x_o}\rvert,
\end{eqnarray}
where $x_o$ is a integral constant. For $\epsilon\neq\frac{1}{4}$, Eq.~(\ref{sec3bb-u-sol-tr}) can be integrated out to be
\begin{eqnarray}
\label{sec3bb-u-sol-tr-b}
\mu_{+}\log\lvert\zeta-\mu_{+}\rvert-\mu_{-}\log\lvert\zeta-\mu_{-}\rvert=-\left(\mu_{+}-\mu_{-}\right)\log\lvert\tfrac{x}{x_o}\rvert.
\end{eqnarray}
This equation expresses $\zeta(x)$ as a implicit function of $x$. With this equation, $\tfrac{x}{x_o}$ can be solved as
\begin{eqnarray}
\label{sec3bb-u-sol-tr-x}
\tfrac{x}{x_o}=\pm\tfrac{1}{\zeta-\mu_{+}}\left(\tfrac{\zeta-\mu_{-}}{\zeta-\mu_{+}}\right)^{\tfrac{\mu_{-}}{\mu_{+}-\mu_{-}}},
\end{eqnarray}
where $\zeta$ is taking the value as
\begin{eqnarray}
\label{sec3bb-u-sol-tr-xz}
\mu_{+}\leq\zeta<\infty.
\end{eqnarray}
In this case, the potential $U(x)$ has a complicated formulation. The solution of Eq.~(\ref{sec3b-osc-qm-sta-v}) can be analyzed with the qualitative method. Using the variable $\zeta$, Eq.~(\ref{sec3b-osc-qm-sta-v}) can be rewritten as
\begin{eqnarray}
\label{sec3bb-osc-qm-sta-v}
-\frac{\hbar^2}{2m}\frac{d\zeta}{dx}\frac{d}{d\zeta}\left(\frac{d\zeta}{dx}\frac{d\psi}{d\zeta}\right)+U(\zeta)\psi(\zeta)-E\psi(\zeta)=0,
\end{eqnarray}
which can be expressed as
\begin{eqnarray}
\label{sec3bb-osc-qm-sta-vs}
-\frac{\hbar^2}{2m}\frac{1}{x^2_o}\frac{d}{d\zeta}\left[b(\zeta)\frac{d\psi}{d\zeta}\right]+\frac{U(\zeta)}{b(\zeta)}\psi(\zeta)-\frac{E}{b(\zeta)}\psi(\zeta)=0,
\end{eqnarray}
where
\begin{eqnarray}
\label{sec3bb-osc-qm-sta-vsb}
b(\zeta)=\zeta^{-1}\left(\zeta-\mu_{+}\right)^{2+\tfrac{\mu_{-}}{\mu_{+}-\mu_{-}}}\left(\zeta-\mu_{-}\right)^{1-\tfrac{\mu_{-}}{\mu_{+}-\mu_{-}}}.
\end{eqnarray}
Notice that $\frac{x}{x_o}$ has two branch solutions in Eq.~(\ref{sec3bb-u-sol-tr-x}). However, these two branches produce the same equation~(\ref{sec3bb-osc-qm-sta-vs}).Therefore, it is enough to analyze the single equation~(\ref{sec3bb-osc-qm-sta-vs}).

From Eq.~(\ref{sec3bb-osc-qm-sta-vs}), we obtain
\begin{eqnarray}
\label{sec3bb-osc-qm-sta-vsi}
E\int^{\infty}_{\mu_{+}}\frac{\psi^2}{b(\zeta)}d\zeta=-\left[\frac{\hbar^2}{2m}\frac{b(\zeta)}{x^2_o}\psi\frac{d\psi}{d\zeta}\right]^{\infty}_{\mu_{+}}+\frac{\hbar^2}{2m}\frac{1}{x^2_o}\int^{\infty}_{\mu_{+}}b(\zeta)\left(\frac{d\psi}{d\zeta}\right)^2d\zeta+\int^{\infty}_{\mu_{+}}\frac{U(\zeta)}{b(\zeta)}\psi^2d\zeta.~~
\end{eqnarray}
For the value of $\zeta$ in Eq.~(\ref{sec3bb-u-sol-tr-xz}), we have
\begin{eqnarray}
\label{sec3bb-osc-qm-sta-bb}
b(\zeta)\geq 0,~~U(\zeta)\geq 0.
\end{eqnarray}
Therefore, according to the Sturm-Liouville theorem, $E\geq 0$ can be implemented by the boundary conditions
\begin{eqnarray}
\label{sec3bb-osc-qm-sta-vsc}
\left[b(\zeta)\psi\frac{d\psi}{d\zeta}\right]^{\infty}_{\mu_{+}}=0.
\end{eqnarray}

In summary, we have discussed two types of solutions of Eq.~(\ref{sec3b-osc-qm-sta-a}) and~(\ref{sec3b-osc-qm-sta-v}). The first type of solutions can be derived in the exact formulations, in which the eigenenergies are shown to be positive. The second type of solution has been treated with the Sturm-Liouville theory, in which the positive energy can be guaranteed by appropriate boundary conditions. 

\section{Higher Dimensional Space}\label{sec4}

We have dealt with higher derivative theories in one dimensional space in the last two sections. In this section, we discuss the higher derivative Lagrangian in the higher dimensional space. We consider the Lagrangian
\begin{eqnarray}
\label{sec4-lag}
L=-\frac{m\epsilon}{2\omega^2}\ddot{x}_{i}\ddot{x}_{i}+\frac{1}{2}m\dot{x}_{i}\dot{x}_{i}-V(r),
\end{eqnarray}
where $i=1,2,3,\cdots d$, and $d$ is the dimensional number of space. $r$ is defined as
\begin{eqnarray}
\label{sec4-lag-r}
r=\sqrt{x^2_1+x^2_2+\cdots+x^2_d}. 
\end{eqnarray}
In Eq.~(\ref{sec4-lag}), we have used Einstein's summation conventions.

We employ Caratheodory's method to analyze mechanical systems in the higher dimensional space. For the Lagrangian~(\ref{sec4-lag}), we construct the surface terms as follows
\begin{eqnarray}
\label{sec4-cara}
-\frac{m\epsilon}{2\omega^2}\ddot{x}_{i}\ddot{x}_{i}+\frac{1}{2}m\dot{x}_{i}\dot{x}_{i}-V(r)=\frac{dF(\mathbf{x},t)}{dt}+\frac{d}{dt}\left(\dot{x}_{i}f_{i}(\mathbf{x},t)\right),
\end{eqnarray}
where $\mathbf{x}$ stands for the higher dimensional coordinates $x_i$. Eq.~(\ref{sec4-cara}) can be expanded as
\begin{eqnarray}
\label{sec4-cara-com}
-\frac{m\epsilon}{2\omega^2}\ddot{x}_{i}\ddot{x}_{i}+\frac{1}{2}m\dot{x}_{i}\dot{x}_{i}-V(r)=
\frac{\partial F}{\partial t}+\dot{x}_i\frac{\partial F}{\partial x_i}+\ddot{x}_{i}f_i
+\dot{x}_{i}\left(\frac{\partial f_i}{\partial t}+\dot{x}_j\frac{\partial f_i}{\partial x_j}\right).
\end{eqnarray}
Similar to Eqs.~(\ref{sec3-osc-cara-2nd-dx}) and (\ref{sec3-osc-cara-2nd-2dx}), we obtain from Eq.~(\ref{sec4-cara-com})
\begin{eqnarray}
\label{sec4-com-sec}
f_i=-\frac{m\epsilon}{\omega^2}\ddot{x}_{i}&=&\frac{\partial L}{\partial \ddot{x}_i},\\
\label{sec4-com-fir}
\frac{\partial F}{\partial x_i}
+\frac{\partial f_i}{\partial t}+\dot{x}_j\left(\frac{\partial f_i}{\partial x_j}+\frac{\partial f_j}{\partial x_i}\right)=m\dot{x}_{i}&=&\frac{\partial L}{\partial \dot{x}_i}.
\end{eqnarray}
We define
\begin{eqnarray}
\label{sec4-def-v}
\dot{x}_{i}=v_i(\mathbf{x},t),\\
\label{sec4-def-vs}
F=S-v_{i}f_i.
\end{eqnarray}
Then we have
\begin{eqnarray}
\label{sec4-def-f}
\ddot{x}_{i}=\frac{d v_i}{dt}=\frac{\partial v_i}{\partial t}+\dot{x}_{j}\frac{\partial v_i}{\partial x_j},
\end{eqnarray}
which is 
\begin{eqnarray}
\label{sec4-def-f-a}
a_i=\ddot{x}_{i}=\frac{\partial v_i}{\partial t}+v_j\frac{\partial v_i}{\partial x_j}.
\end{eqnarray}
This is Euler's formula in Fluid mechanics. With these equations, Eqs.~(\ref{sec4-com-sec}) and (\ref{sec4-com-fir}) are computed as
\begin{eqnarray}
\label{sec4-com-sec-re}
f_i&=&-\frac{m\epsilon}{\omega^2}\left(\frac{\partial v_i}{\partial t}+v_j\frac{\partial v_i}{\partial x_j}\right),\\
\label{sec4-com-fir-re}
mv_{i}&=&\frac{\partial S}{\partial x_i}
+\frac{\partial f_i}{\partial t}+v_j\frac{\partial f_i}{\partial x_j}-f_j\frac{\partial v_j}{\partial x_i}.
\end{eqnarray}
With its definition in Eq.~(\ref{sec4-com-sec-re}), $f_i$ can be eliminated from Eq.~(\ref{sec4-com-fir-re}), then we obtain
\begin{eqnarray}
\label{sec4-cara-v}
\frac{\partial S}{\partial x_i}=mv_i&+&\frac{m\epsilon}{\omega^2}\left(\frac{\partial^2 v_i}{\partial t^2}+2v_{k}\frac{\partial^2 v_i}{\partial t\partial x_k}+v_j v_k\frac{\partial^2 v_i}{\partial x_j\partial x_k}\right)\\
&+&\frac{m\epsilon}{\omega^2}\left(\frac{\partial v_j}{\partial t}+v_{k}\frac{\partial v_j}{\partial x_k}\right)s_{ij},\nonumber
\end{eqnarray}
where
\begin{eqnarray}
\label{sec4-cara-v-s}
s_{ij}=\frac{\partial v_i}{\partial x_j}-\frac{\partial v_j}{\partial x_i}.
\end{eqnarray}
In contrast with Eq.~(\ref{sec3-cara-2nd-dx-sol-f}) in one dimensional space, there is additionally a rotational term $s_{ij}$ in the higher dimensional case. Eq.~(\ref{sec4-cara-com}) can be computed as
\begin{eqnarray}
\label{sec4-cara-com-b}
\frac{\partial S}{\partial t}+v_i\frac{\partial S}{\partial x_i}=-\frac{m\epsilon}{2\omega^2}\left(\frac{\partial v_i}{\partial t}+v_j\frac{\partial v_i}{\partial x_j}\right) \left(\frac{\partial v_i}{\partial t}+v_k\frac{\partial v_i}{\partial x_k}\right)+\frac{1}{2}mv_{i}v_{i}-V(r),
\end{eqnarray}
where $v_i$ is determined by Eq.~(\ref{sec4-cara-v}). Eqs.~(\ref{sec4-cara-v}) and (\ref{sec4-cara-com-b}) are the derived novel formulations of Hamilton-Jacobi equations in the higher dimensional space. These equations employ the velocity field $v_i$ as the important variable instead of canonical variables.

In the below, we begin to work out several equivalent deformations of Eq.~(\ref{sec4-cara-com-b}). At first, we can rewrite Eq.~(\ref{sec4-cara-com-b}) as
\begin{eqnarray}
\label{sec4-cara-com-re}
\frac{\partial S}{\partial t}+\frac{1}{2m}\frac{\partial S}{\partial x_i}\frac{\partial S}{\partial x_i}+V(r)&=&\frac{1}{2m}\left(mv_i-\frac{\partial S}{\partial x_i}\right)\left(mv_i-\frac{\partial S}{\partial x_i}\right)\\
&-&\frac{m\epsilon}{2\omega^2}\left(\frac{\partial v_i}{\partial t}+v_j\frac{\partial v_i}{\partial x_j}\right) \left(\frac{\partial v_i}{\partial t}+v_k\frac{\partial v_i}{\partial x_k}\right).\nonumber
\end{eqnarray}
We make the definition
\begin{eqnarray}
\label{sec4-osc-v-def}
u_i=v_i-\frac{1}{m}\frac{\partial S}{\partial x_i},
\end{eqnarray}
then $v_i$ is expressed as
\begin{eqnarray}
\label{sec4-osc-u-def}
v_i=u_i+\frac{1}{m}\frac{\partial S}{\partial x_i}.
\end{eqnarray}
We also define
\begin{eqnarray}
\label{sec4-cara-v-w}
K_i=-\frac{\epsilon}{\omega^2}\left(\frac{\partial^2 v_i}{\partial t^2}+2v_{k}\frac{\partial^2 v_i}{\partial t\partial x_k}+v_j v_k\frac{\partial^2 v_i}{\partial x_j\partial x_k}\right)
-\frac{\epsilon}{\omega^2}\left(\frac{\partial v_j}{\partial t}+v_{k}\frac{\partial v_j}{\partial x_k}\right)s_{ij},
\end{eqnarray}
then we obtain from Eq.~(\ref{sec4-cara-v})
\begin{eqnarray}
\label{sec4-osc-u-w}
u_i-K_i=0.
\end{eqnarray}
Using Eq.~(\ref{sec4-osc-u-def}) and the Leibniz differential rules, we have the identity
\begin{eqnarray}
\label{sec4-v-u-w}
\frac{\partial v_i}{\partial t}+v_{j}\frac{\partial v_i}{\partial x_j}&=&\frac{\partial}{\partial t}\left(u_i+\frac{1}{m}\frac{\partial S}{\partial x_i}\right)+\left(u_j+\frac{1}{m}\frac{\partial S}{\partial x_j}\right)\frac{\partial}{\partial x_j}\left(u_i+\frac{1}{m}\frac{\partial S}{\partial x_i}\right)\nonumber\\
&=&\frac{\partial u_i}{\partial t}+u_{j}\frac{\partial u_i}{\partial x_j}+\frac{1}{m}\frac{\partial}{\partial x_i}\left(M+u_j\frac{\partial S}{\partial x_j}\right)
+\frac{1}{m}\frac{\partial S}{\partial x_j}\left(\frac{\partial u_i}{\partial x_j}-\frac{\partial u_j}{\partial x_i}\right),
\end{eqnarray}
where $M$ is defined as
\begin{eqnarray}
\label{sec4-v-u-w-m}
M=\frac{\partial S}{\partial t}
+\frac{1}{2m}\frac{\partial S}{\partial x_i}\frac{\partial S}{\partial x_i}.
\end{eqnarray}
With the consideration of Eq.~(\ref{sec4-osc-u-w}), Eq.~(\ref{sec4-v-u-w}) is rewritten as
\begin{eqnarray}
\label{sec4-v-u-w-re}
\frac{\partial v_i}{\partial t}+v_{j}\frac{\partial v_i}{\partial x_j}=
\frac{\partial K_i}{\partial t}+K_{j}\frac{\partial K_i}{\partial x_j}+\frac{1}{m}\frac{\partial}{\partial x_i}\left(M+K_j\frac{\partial S}{\partial x_j}\right)
+\frac{1}{m}K_{ij}\frac{\partial S}{\partial x_j},
\end{eqnarray}
where
\begin{eqnarray}
\label{sec4-v-u-w-w}
K_{ij}=\frac{\partial K_i}{\partial x_j}-\frac{\partial K_j}{\partial x_i}.
\end{eqnarray}
Therefore, compared with the equation in section~\ref{sec3}, Eq.~(\ref{sec4-v-u-w-re}) has an additional term associated with the rotational quantity $K_{ij}$ in the higher dimensional space. With the aforementioned relations, Eq.~(\ref{sec4-cara-com-re}) is finally recast as
\begin{eqnarray}
\label{sec4-cara-com-re-k}
\frac{\partial S}{\partial t}+\frac{1}{2m}\frac{\partial S}{\partial x_i}\frac{\partial S}{\partial x_i}+V(r)=\frac{m}{2} K_i K_i-\frac{m\epsilon}{2\omega^2}\mathcal{A}_i\mathcal{A}_i,
\end{eqnarray}
where $\mathcal{A}_i$ is
\begin{eqnarray}
\label{sec4-v-u-w-re-a}
\mathcal{A}_i=
\frac{\partial K_i}{\partial t}+K_{j}\frac{\partial K_i}{\partial x_j}+\frac{1}{m}\frac{\partial}{\partial x_i}\left(M+K_j\frac{\partial S}{\partial x_j}\right)
+\frac{1}{m}K_{ij}\frac{\partial S}{\partial x_j}.
\end{eqnarray}
Eq.~(\ref{sec4-cara-com-re-k}) is also
\begin{eqnarray}
\label{sec4-cara-com-re-b}
M+V(r)=\frac{m}{2} K_i K_i-\frac{m\epsilon}{2\omega^2}\mathcal{A}_i\mathcal{A}_i,
\end{eqnarray}
where $M$ and $K_i$ are defined by Eqs.~(\ref{sec4-v-u-w-m}) and (\ref{sec4-cara-v-w}). The elemental variables in Eq.~(\ref{sec4-cara-com-re-b}) are $S$ and $v_i$.

Similar to discussions in subsection~\ref{sec3b}, we begin to search for Schr$\ddot{\mathrm{o}}$dinger-type equations associated with Eqs.~(\ref{sec4-cara-com-re-b}) and (\ref{sec4-cara-v}) in the classical limit $\hbar\rightarrow 0$. The correspondences in Eq.~(\ref{sec3b-osc-v}) can be generalized to the present situation. We make the correspondences
\begin{eqnarray}
\label{sec4-osc-v}
\frac{\partial S}{\partial t}\rightarrow \hbar\mathbf{Im}\left(\frac{1}{\Psi}\frac{\partial \Psi}{\partial t}\right),~~\frac{\partial S}{\partial x_i}\rightarrow \hbar\mathbf{Im}\left(\frac{1}{\Psi}\frac{\partial \Psi}{\partial x_i}\right),
\end{eqnarray}
and the correspondence
\begin{eqnarray}
\label{sec4-osc-vv}
\frac{\partial S}{\partial x_i}\frac{\partial S}{\partial x_i}\rightarrow -\hbar^2\mathbf{Re}\left(\frac{1}{\Psi}\frac{\partial^2 \Psi}{\partial x_i\partial x_i}\right).
\end{eqnarray}

Using the correspondences in Eqs.~(\ref{sec4-osc-v}) and (\ref{sec4-osc-vv}), we obtain from Eq.~(\ref{sec4-cara-com-re-b})
\begin{eqnarray}
\label{sec4-osc-qm}
\mathbf{M}+V(r)
=\frac{1}{2}m K_i K_i-\frac{m\epsilon}{2\omega^2}\mathbf{A}_i\mathbf{A}_i,
\end{eqnarray}
where
\begin{eqnarray}
\label{sec4-osc-qm-a}
\mathbf{A}_i&=&
\frac{\partial K_i}{\partial t}+K_{j}\frac{\partial K_i}{\partial x_j}+\frac{1}{m}\frac{\partial}{\partial x_i}\left[\mathbf{M}+\hbar K_j\mathbf{Im}(\tfrac{1}{\Psi}\tfrac{\partial\Psi}{\partial x_j})\right]
+\hbar K_{ij}\mathbf{Im}(\tfrac{1}{\Psi}\tfrac{\partial\Psi}{\partial x_j}),\\
\label{sec4-osc-qm-m}
\mathbf{M}&=&\hbar\mathbf{Im}(\tfrac{1}{\Psi}\tfrac{\partial\Psi}{\partial t})
-\tfrac{\hbar^2}{2m}\mathbf{Re}\left(\tfrac{1}{\Psi}\tfrac{\partial^2 \Psi}{\partial x_i\partial x_i}\right).
\end{eqnarray}
The formulations of $K_i$ and $K_{ij}$ remain unchanged as their definitions in Eqs.~(\ref{sec4-cara-v-w}) and (\ref{sec4-v-u-w-w}). Eq.~(\ref{sec4-cara-v}) is transformed to be the following formulation
\begin{eqnarray}
\label{sec4-cara-v-qm}
\hbar\mathbf{Im}\left(\frac{1}{\Psi}\frac{\partial\Psi}{\partial x_i}\right)=mv_i&+&\frac{m\epsilon}{\omega^2}\left(\frac{\partial^2 v_i}{\partial t^2}+2v_{k}\frac{\partial^2 v_i}{\partial t\partial x_k}+v_j v_k\frac{\partial^2 v_i}{\partial x_j\partial x_k}\right)\\
&+&\frac{m\epsilon}{\omega^2}\left(\frac{\partial v_j}{\partial t}+v_{k}\frac{\partial v_j}{\partial x_k}\right)s_{ij},\nonumber
\end{eqnarray}
where $s_{ij}$ is defined by Eq.~(\ref{sec4-cara-v-s}). In the higher dimensional space, we also have the continuity equation
\begin{eqnarray}
\label{sec4-con-p}
\frac{\partial \rho}{\partial t}+\frac{\partial}{\partial x_i}(\rho v_i)=0.
\end{eqnarray}
With $\rho=\Psi\Psi^{*}$, it is represented as
\begin{eqnarray}
\label{sec4-con-p-r}
\frac{\partial}{\partial t}(\Psi\Psi^{*})+\frac{\hbar}{m}\frac{\partial}{\partial x_i}\left[(\Psi\Psi^{*})v_i\right]=0.
\end{eqnarray}
Eqs.~(\ref{sec4-osc-qm}), (\ref{sec4-cara-v-qm}) and (\ref{sec4-con-p-r}) are the derived Schr$\ddot{\mathrm{o}}$dinger-type equations in the higher dimensional space, which generate the Hamilton-Jacobi equations in the classical limit $\hbar\rightarrow 0$. These three equations are complete systems to determine the wave function $\Psi$ and the velocity field $v_i$. When $\epsilon=0$, these equations are degenerated into the conventional Schr$\ddot{\mathrm{o}}$dinger equation. In comparison with the one dimensional case in subsection~\ref{sec3b}, there are new contributions from the rotational term $K_{ij}$. 

In the higher dimensional space, the solutions of Schr$\ddot{\mathrm{o}}$dinger-type equations are more complicated. As a preliminary analysis, we shall deal with the simplest case in the below. We consider the stationary solutions of spherical symmetry as follows
\begin{eqnarray}
\label{sec4-sol-u}
\Psi(t,\mathbf{x})=\psi(r)e^{\frac{i}{\hbar}\theta(r)}e^{-\frac{i}{\hbar}Et},~~
v_i(t,\mathbf{x})=\chi(r)\frac{x_i}{r}.
\end{eqnarray}
$K_i$ is determined by $v_i$ in Eq.~(\ref{sec4-cara-v-w}), then we obtain
\begin{eqnarray}
\label{sec4-sol-u-w}
K_i=K(r)\frac{x_i}{r},~~K(r)=-\frac{\epsilon}{\omega^2}\chi^2\frac{d^2\chi}{dr^2}.
\end{eqnarray}
The continuity equation~(\ref{sec4-con-p-r}) becomes
\begin{eqnarray}
\label{sec4-con-p-r-chi}
\frac{d}{dr}\left(\psi^{2}\chi\right)+\frac{d-1}{r}\left(\psi^{2}\chi\right)=0,
\end{eqnarray}
which solution is
\begin{eqnarray}
\label{sec4-con-p-r-chi-sol}
\chi(r)=\frac{\delta}{\psi(r)^{2}}r^{1-d}.
\end{eqnarray}
Here $\delta$ is a integral constant. Eq.~(\ref{sec4-osc-qm}) is simplified to
\begin{eqnarray}
\label{sec4-osc-qm-sta}
\mathbf{M}+V(r)
=\frac{m}{2}K^2-\frac{m\epsilon}{2\omega^2}\mathbf{A}^2,
\end{eqnarray}
where $\mathbf{A}(r)$ and $\mathbf{M}(r)$ are computed as
\begin{eqnarray}
\label{sec4-osc-qm-sta-a}
\mathbf{A}(r)&=&K\frac{dK}{dr}+\frac{1}{m}\left(\frac{d\mathbf{M}}{dr}+K\frac{d^2\theta}{dr^2}+\frac{dK}{dr}\frac{d\theta}{dr}\right),\\
\label{sec4-osc-qm-sta-m}
\mathbf{M}(r)&=&-E-\frac{\hbar^2}{2m}\left(\frac{1}{\psi}\frac{d^2\psi}{d r^2}+\frac{d-1}{r}\frac{1}{\psi}\frac{d\psi}{dr}\right)+\frac{1}{2m}\left(\frac{d\theta}{dr}\right)^2.
\end{eqnarray}
Eq.~(\ref{sec4-cara-v-qm}) turns into
\begin{eqnarray}
\label{sec3b-dx-sol-f-r}
\frac{d\theta}{dr}=m\chi+\frac{m\epsilon}{\omega^2}\chi^2\frac{d^2\chi}{dr^2}.
\end{eqnarray}
This is a single equation to express $\theta(r)$ in terms of $\chi(r)$. For the integral constant $\delta\neq 0$ in Eq.~(\ref{sec4-con-p-r-chi-sol}), it is difficult to solve Eq.~(\ref{sec4-osc-qm-sta}). We consider the simpler case $\delta= 0$. For $\delta= 0$, we obtain from Eq.~(\ref{sec4-con-p-r-chi-sol})
\begin{eqnarray}
\label{sec4-con-p-r-chi-d}
\chi(r)=0.
\end{eqnarray}
In this situation, Eq.~(\ref{sec4-osc-qm-sta}) is tremendously simplified. Similar to Eq.~(\ref{sec3b-osc-qm-sta}), it is equivalent to the following two equations
\begin{eqnarray}
\label{sec4-osc-qm-sta-a}
-V(r)+U(r)
-\frac{\epsilon}{2m\omega^2}\left(\frac{dU}{dr}\right)^2&=&0,\\
\label{sec4-osc-qm-sta-v}
-\frac{\hbar^2}{2m}\left(\frac{1}{\psi}\frac{d^2\psi}{d r^2}+\frac{d-1}{r}\frac{1}{\psi}\frac{d\psi}{dr}\right)&=&E-U(r).
\end{eqnarray}
Eq.~(\ref{sec4-osc-qm-sta-v}) is the stationary Schr$\ddot{\mathrm{o}}$dinger equation of spherical symmetry in the higher dimensional space, which potential is determined by Eq.~(\ref{sec4-osc-qm-sta-a}). Eq.~(\ref{sec4-osc-qm-sta-a}) is similar to Eq.~(\ref{sec3b-osc-qm-sta-a}) in subsection~\ref{sec3b}.  The discussions about the free particle potential and the harmonic oscillator potential follow from that in subsections~\ref{sec3ba} and~\ref{sec3bb}.

As an example and for practical applications in three dimensional space, we consider the attractive Coulomb potential
\begin{eqnarray}
\label{sec4-col}
V(r)=-\frac{\lambda^2}{r},
\end{eqnarray}
where $\lambda$ is the coupling constant. At the first step, we seek the solution of $U(r)$. Eq.~(\ref{sec4-osc-qm-sta-a}) turns into
\begin{eqnarray}
\label{sec4-col-qm}
\frac{\lambda^2}{r}+U(r)
-\frac{\epsilon}{2m\omega^2}\left(\frac{dU}{dr}\right)^2=0.
\end{eqnarray}
This equation has two kinds of series solutions. The first kind of series solution is
\begin{eqnarray}
\label{sec4-col-qm-sa}
U(r)=-\tfrac{\lambda^2}{r_0}&+&\tfrac{2\omega\lambda\sqrt{2mr_0}}{\sqrt{\epsilon}}\biggl[\left(\tfrac{r}{r_0}\right)^{\tfrac{1}{2}}+\tfrac{1}{6}\left(\tfrac{r}{r_0}\right)^{\tfrac{3}{2}}+\tfrac{1}{40}\left(\tfrac{r}{r_0}\right)^{\tfrac{5}{2}}\biggr]\\
&+&\tfrac{m\omega^2r^2_0}{\epsilon}\left[\left(\tfrac{r}{r_0}\right)^2+\tfrac{2}{9}\left(\tfrac{r}{r_0}\right)^3\right]+\cdots,\nonumber
\end{eqnarray}
where $r_0$ is the integral constant. This series solution behaves well defined at the short distance. The second kind of series solution is the naive expansion about $\epsilon$, which yields
\begin{eqnarray}
\label{sec4-col-qm-sb}
U(r)=-\frac{\lambda^2}{r}+\frac{1}{2}\frac{\epsilon\lambda^2}{m\omega^2}\frac{\lambda^2}{r^4}-2\left(\frac{\epsilon\lambda^2}{m\omega^2}\right)^2\frac{\lambda^2}{r^7}+16\left(\frac{\epsilon\lambda^2}{m\omega^2}\right)^3\frac{\lambda^2}{r^{10}}+\cdots.
\end{eqnarray}
This series solution behaves well defined at the large distance, which converges to the Coulomb potential when $r$ is enough large, but it is divergent at the short distance. Aside from the aforementioned series solutions, the exact solution of $U(r)$ can be found as follows. With the definition
\begin{eqnarray}
\label{sec4-col-de}
U(r)=\mathcal{U}(r)^2-\frac{\lambda^2}{r},
\end{eqnarray}
Eq.~(\ref{sec4-col-qm}) can be recast as
\begin{eqnarray}
\label{sec4-col-re}
\frac{d\mathcal{U}}{dr}=\frac{\omega\sqrt{m}}{\sqrt{2\epsilon}}-\frac{\lambda^2}{2r^2}\frac{1}{\mathcal{U}}.
\end{eqnarray}
Using the transformations
\begin{eqnarray}
\label{sec4-col-tr}
\mathcal{U}&=&\left(\tfrac{\omega\lambda^2\sqrt{2m}}{\sqrt{\epsilon}}\right)^{\frac{2}{3}}\tfrac{1}{\mathscr{U}(\zeta)}\tfrac{d\mathscr{U}(\zeta)}{d\zeta}+\tfrac{\omega\sqrt{m}}{\sqrt{2\epsilon}}r,\\
\zeta&=&\left(\tfrac{\sqrt{\epsilon}}{\omega\lambda^2\sqrt{2m}}\right)^{\frac{1}{3}}\left[\left(\mathcal{U}(r)-\tfrac{\omega\sqrt{m}}{\sqrt{2\epsilon}}r\right)^2-\tfrac{\lambda^2}{r}\right],
\end{eqnarray}
we can derive from Eq.~(\ref{sec4-col-re})
\begin{eqnarray}
\label{sec4-col-re-b}
\frac{d^2\mathscr{U}}{d\zeta^2}-\zeta\cdot\mathscr{U}(\zeta)=0.
\end{eqnarray}
This equation can be solved by the Airy functions
\begin{eqnarray}
\label{sec4-col-re-b-sol}
\mathscr{U}(\zeta)=c_1 \mathrm{AiryAi}(\zeta)+c_2 \mathrm{AiryBi}(\zeta).
\end{eqnarray}
Finally $U(r)$ is determined by the implicit function
\begin{eqnarray}
\label{sec4-col-re-b-sol-b}
\sqrt{U(r)+\tfrac{\lambda^2}{r}}-\tfrac{\omega\sqrt{m}}{\sqrt{2\epsilon}}r&=&\left(\tfrac{\omega\lambda^2\sqrt{2m}}{\sqrt{\epsilon}}\right)^{\frac{2}{3}}\tfrac{c_1 \mathrm{AiryAi}(1,\zeta)+c_2 \mathrm{AiryBi}(1,\zeta)}{c_1 \mathrm{AiryAi}(\zeta)+c_2 \mathrm{AiryBi}(\zeta)},\\
\left(\tfrac{\omega\lambda^2\sqrt{2m}}{\sqrt{\epsilon}}\right)^{\frac{1}{3}}\zeta(r)&=&\left(\sqrt{U(r)+\tfrac{\lambda^2}{r}}-\tfrac{\omega\sqrt{m}}{\sqrt{2\epsilon}}r\right)^2-\tfrac{\lambda^2}{r}.
\end{eqnarray}
Because of the complicated series formulations of $U(r)$, we could merely be able to derive qualitative or numerical results for the Coulomb potential.

\section{More Discussions and Conclusions}\label{sec5}

With the help of Caratheodory's equivalent Lagrangian method, we have shown that there exist novel formulations of Hamilton-Jacobi equations for higher derivative theories in sections~\ref{sec3} and~\ref{sec4}. Because the velocity field plays a critical role in these novel formulations, we suggest their names as the velocity field formalism of Hamilton-Jacobi equations. These novel Hamilton-Jacobi equations are different from the canonical ones. Moreover, we analyzed the classical solutions of these equations and presented their quantum mechanical correspondences in subsections~\ref{sec3a}, \ref{sec3b} and~\ref{sec4} respectively. Using the Sturm-Liouville theory, we have shown that the unbounded negative energy problem could be avoided in the derived Schr$\ddot{\mathrm{o}}$dinger-type equations.

The induced Schr$\ddot{\mathrm{o}}$dinger-type equations turn out to be nonlinear systems. The nonlinear time evolutions of wave functions make it difficult to define quantum entropy~\cite{Peres:1989zz,Weinberg:1989an} and cause other theoretical issues~\cite{Gisin:1989sx,Polchinski:1990py,Abrams:1998vz}, which require furthermore detailed considerations.

In Refs.~\cite{Weinberg:1989us,Weinberg:1989cm,Bollinger:1989zz,Walsworth:1990zz,Majumder:1990zz}, nonlinear quantum mechanical models are proposed and experimental tests are performed to restrict the nonlinear new parameters. The new parameters in higher derivative theories could be constrained by a similar way.

We should mention that the Schr$\ddot{\mathrm{o}}$dinger-type equations in subsections~\ref{sec3b} and~\ref{sec4} are derived with the help of intuitive procedures, which are not able to determine the Schr$\ddot{\mathrm{o}}$dinger-type equations uniquely. There exist other Schr$\ddot{\mathrm{o}}$dinger-type equations, which have different formulations but produce the same classical limits. It is a challenge to find additional guiding principles to restrict the final formulations of Schr$\ddot{\mathrm{o}}$dinger-type equations.

We have focused on the mechanical models in the present discussions. We can also derive novel Hamilton-Jacobi equations for field theories, and discuss the corresponding quantum theoretical models. It remains as an attempt to work out some possible applications on higher derivative gravitational theories, scalar field theories and gauge theories.

\bibliographystyle{utphys}

\bibliography{Highderivative}

\end{document}